\documentclass[aps,prb,preprint,superscriptaddress,amsmath,showpacs,amssymb]{revtex4}
\usepackage{graphicx}

\begin{document}

\title{Discrete easy-axis tilting in Mn$_{12}$-acetate, as determined by EPR: implications for the magnetic quantum tunneling
mechanism}
\author{S. Takahashi}
\affiliation{Department of Physics, University of Florida,
Gainesville, FL 32611,USA}
\author{R. S. Edwards}
\affiliation{Department of Physics, University of Florida,
Gainesville, FL 32611,USA}
\author{J. M. North}
\affiliation{Department of Chemistry and National High Magnetic
Field Laboratory, Tallahassee, FL 32310, USA}
\author{S. Hill}
\email[corresponding author, Email:]{hill@phys.ufl.edu}
\affiliation{Department of Physics, University of Florida,
Gainesville, FL 32611,USA}
\author{N. S. Dalal}
\affiliation{Department of Chemistry and National High Magnetic
Field Laboratory, Tallahassee, FL 32310, USA}

\date{\today}

\begin{abstract}
The variation with microwave frequency and temperature of previously
reported anomalous peaks in the EPR spectra of Mn$_{12}$-acetate,
under large transverse fields, reveals that the molecular easy
magnetization axes are tilted with respect to the global symmetry
direction. More importantly, on the basis of the angle-dependence of
fine structures observed in the EPR spectra we infer that the tilt
distribution must be discrete, as was previously suspected from
studies which demonstrated the presence of a locally varying rhombic
anisotropy [S. Hill {\em et al.}, Phys. Rev. Lett. {\bf 90}, 217204
(2003)]. The tilts are confined to two orthogonal planes, and the
distribution extends up to $\sim 1.7^\circ$ degrees away from the
the global easy ($z$-) axis. We ascribe the tilting to the
hydrogen-bonding effect associated with the disordered acetic acid
solvent molecules. The effect is considerably larger than deduced
from x-ray diffraction analyses. These data constitute the
sought-after evidence for the presence of transverse fields in
Mn$_{12}$-acetate, and provide a possible explanation for the lack
of selection rules in the resonant quantum tunneling behavior seen
in low-temperature hysteresis experiments for this $S = 10$ system.
\end{abstract}


\maketitle

\section{Introduction}

Since the discovery of magnetic quantum tunneling (MQT) in
[Mn$_{12}$O$_{12}$(CH$_3$COO)$_{16}$(H$_2$O)$_4$]$\cdot$
2CH$_3$COOH$\cdot4$H$_2$O
(Mn$_{12}$-ac),\cite{Sessoli93a,Sessoli93b,Sessoli95,Novak95,Friedman96,Thomas96,MRS00,Angewandte03}
single molecule magnets (SMMs) have become the focus of
considerable experimental and theoretical interest due to their
novel quantum properties \cite{MRS00,Angewandte03} and possible
future use in quantum computational
devices.\cite{LossNature01,HillScience} Mn$_{12}$-ac remains the
most widely studied SMM due to its large spin ground state ($S =
10$, see refs [\onlinecite{Friedman96,Thomas96,
MRS00,Angewandte03}]), together with a considerable easy-axis
magneto-crystalline anisotropy.\cite{MRS00,Angewandte03} These
combined factors result in a sizeable kinetic barrier against spin
reversal at the molecular level, leading to slow magnetization
relaxation and hysteresis (bistability) at low temperatures
(below~$\sim$3~K).\cite{Sessoli95,Novak95} When a DC magnetic
field is applied parallel to the easy-axis of a single crystal of
Mn$_{12}$-ac, sharp steps are observed in its hysteresis loops at
well defined field
strengths.\cite{Friedman96,Thomas96,MRS00,Angewandte03,Perenboom}
The enhanced magnetization relaxation at these steps is the result
of resonant MQT.

While a clearer picture concerning the mechanism of MQT in
Mn$_{12}$-ac is beginning to emerge,
\cite{ChudnPRL01,CorniaPRL02,HillPRL03a,delBarco03} many interesting
problems remain. In particular, current theoretical models assume
the presence of quadratic and quartic transverse crystal-field
interactions in the spin Hamiltonian [$\hat{O}^2_2=\frac{1}{2}(S_+^2
+ S_-^2)$ and $\hat{O}^4_4=\frac{1}{2}(S_+^4 + S_-^4)$], where the
former has been ascribed to solvent disorder.\cite{CorniaPRL02}
However, these interactions, which contain only even powers of the
raising and lowering operators, {\em do not} provide an explanation
for the observation of 'odd' MQT steps in the hysteresis loops, i.e.
tunneling via resonances between levels whose spin projections
($m_s$) differ by an odd integer.\cite{MRS00,Angewandte03} It is
generally recognized that the underlying mechanism {\em must}
involve internal transverse fields,\cite{ChudnPRL01,WernsEPL99} but
their source has yet to be identified. Prokof'ev and
Stamp\cite{Stamp} proposed that hyperfine interactions might provide
the answer; however, the calculated MQT rates were too small.
Subsequently, Garanin and Chudnovsky\cite{ChudnPRL01} suggested that
strains or structural defects in the crystal lattice (dislocations)
could lead to a distribution of tilts of the magnetic easy-axes at
the Mn$_{12}$-ac cluster sites; upon application of an external
magnetic field, such tilts would result in a distribution of
transverse fields, even when the field is applied parallel to the
global symmetry ($z$-) axis of the crystal. While this model was in
qualitative agreement with the fact that experiments revealed
significant distributions in the finite field MQT
rates,\cite{MertesPRL,delBarcoEPL} the measured distributions were
considerably narrower than those predicted by Garanin and
Chudnovsky.\cite{ChudnPRL01}

Although some preliminary spectroscopic data have been reported in
support of dislocations,\cite{Amigo} a more plausible model leading
to easy-axis-tilting was recently suggested by Cornia et
al.\cite{CorniaPRL02} Based on a detailed analysis of existing x-ray
diffraction data and some approximate electronic structure
calculations, they propose that such a local symmetry lowering and
easy-axis-tilting can be ascribed to the presence of the two acetic
acids of crystallization in the unit cell. Experiments in support of
this model were recently published by us,\cite{HillPRL03a} and by
del~Barco et al.\cite{delBarco03} However, evidence for the
easy-axis-tilting was lacking in these investigations, though we
note that tilts have recently been reported for a related Mn$_{12}$
complex in a separate study by del Barco et al.\cite{delBarco04b}
Here we report electron paramagnetic resonance (EPR) measurements on
precisely (in-situ) aligned Mn$_{12}$-ac single crystals under large
applied fields in the transverse direction. At relatively low
frequencies ($<90$~GHz), as the field is rotated away from the hard
plane, simulations show that the EPR intensity should oscillate
between two series of resonances excited from ``even" and ``odd"
spin states (labeled $\alpha$ and $\beta$ respectively, see
following section). The magnetic dipole matrix elements and
transition frequencies associated with these ``even" and ``odd"
resonances are extremely sensitive to the field orientation
(providing $<1^\circ$ resolution), allowing for a precise appraisal
of the Cornia model. Indeed, experiments reveal a significant
overlap of the $\alpha$ and $\beta$ resonances, providing conclusive
evidence for the presence of tilts. However, we find that the
magnitude of the tilts is a factor of $4-6$ larger than predicted by
Cornia et al.,\cite{CorniaPRL02} which should be of both theoretical
and experimental significance in regards to understanding the spin
dynamics of Mn$_{12}$-ac. Our conclusions also provide an
explanation for several previously reported anomalous EPR
transitions (labeled $\beta$-resonances) which cannot be explained
within the widely accepted giant spin ($S=10$) model described
below.\cite{HillPRL98}

We emphasize from the outset that our conclusions are based upon the
angle-dependent behavior of EPR fine structures which have
previously been explained\cite{HillPRL03a} within the context of the
Cornia solvent-disorder picture,\cite{CorniaPRL02} and have been
independently supported by magnetization
measurements.\cite{delBarco03} The easy axis tilting is inferred
from the persistence of EPR peaks for field orientations which are
tilted significantly away from the angles where the intensity should
have vanished if all molecules were aligned. The discrete nature of
the tilting can then be seen from the fact that the different fine
structures exhibit distinct angle dependencies. This behavior is
completely reproducible in Mn$_{12}$-ac samples prepared by
different methods and by different
groups.\cite{delBarco04,HillChakov} Furthermore, our method of
analysis is not sensitive to the intricacies associated with
lineshape analyses, which could be influenced by complex many-body
effects, e.g. magnetic spin-spin interactions.\cite{Werns04}
Finally, we note that all of the unusual solvent-induced anomalies
reported in this paper for Mn$_{12}$-ac are absent in the EPR
spectra for a related high-symmetry ($S_4$) Mn$_{12}$ complex
[Mn$_{12}$O$_{12}$(O$_{2}$CCH$_{2}$Br)$_{16}$(H$_{2}$O)$_{4}\cdot$4CH$_2$Cl$_2$]
which possesses a full compliment of four CH$_2$Cl$_2$ solvent
molecules per Mn$_{12}$.\cite{PetukhovS9}

\section{Background}

The basis of our experiment is illustrated in Fig. 1, which shows
the spin energy level diagram for Mn$_{12}$-ac (Fig.~1a), based on
an exact diagonalization of the standard $S = 10$ spin Hamiltonian
\cite{Angewandte03,CorniaPRL02,HillPRL03a,delBarco03,delBarco04b,HillPRL98,Barra,MirebeauPRL99,Mukhin}
under a transverse magnetic field $B$, and assuming $S_4$
symmetry:

\vspace{6pt}

\noindent{ \hfill  \hfill  \hfill  \hfill  \hfill $\hat H = D\hat
S_z^2 + \mu _B \vec B \cdot
\mathord{\buildrel{\lower3pt\hbox{$\scriptscriptstyle\rightarrow$}}
\over g}  \cdot \hat S + B_4^0\hat{O}_4^0 + B_4^4\hat{O}_4^4,
\hfill\hfill\hfill\hfill\hfill (1)$}

\vspace{6pt}

\noindent{Here, $S_z$ is the projection of the spin operator
$\hat{S}$ along $z$, and $D$ ($<0$) is the uniaxial anisotropy
parameter; the second term is the Zeeman interaction; and the
remaining terms represent higher order crystal field interactions
(the operators $\hat{O}_4^0$ and $\hat{O}_4^4$ have their usual
meaning).\cite{Angewandte03,MirebeauPRL99} We have used acceptable
parameters for the simulations in Fig.~1: $D = -0.455$ cm$^{-1}$,
$B_4^0$ = $-2.0 \times 10^{-5}$ cm$^{-1}$ and $B_4^4$ = $\pm 3.2
\times 10^{-5}$ cm$^{-1}$. We note that this parameter set yields
the optimum agreement with a large body of single crystal EPR
data,\cite{HillPRL03a,HillPolyS9,HillPolyE,note0} including the
present study, which additionally considers the effects of {\em
E}-strain and easy axis tilting. At this stage, we do not
explicitly include lower symmetry interactions in Eq.~1 due to
disorder, e.g. second-order rhombicity, or easy-axis tilting. We
begin by considering a single molecule and the effect of
field-mis-alignment away from the hard plane. We then compare
experimental data with simulations in order to quantify the
easy-axis tilting caused by the solvent-disorder in Mn$_{12}$-ac.
Finally, we consider the rhombic anisotropy associated with this
disorder. We note that an account of the influence of the
solvent-disorder-induced rhombic anisotropy [$E(S_x^2-S_y^2)$],
and its affect on the transverse field EPR fine structure, has
been presented previously by us.\cite{HillPRL03a,HillPolyE} In the
following, we use polar coordinates to parameterize the field
orientation: $\theta$ represents the angle between the applied
field and the global easy ($z$-) axis of the single crystal;
$\phi$ represents the azimuthal angle, i.e. the angle between the
intrinsic hard four-fold ($x$-) $B^4_4$ axis and the projection of
the applied field onto the hard plane.}

In the high-field limit ($g \mu_B${\bf B}~$> |D|S$), and for the
standard EPR geometry (microwave field {\bf H}$_1\perp${\bf B}),
one expects a total of 20 $\Delta m = \pm 1$ EPR transitions
between the 21 ($=2S+1$) spin-states for $S = 10$. For clarity,
only a few of the lowest-lying levels are shown in Fig.~1a for the
case of a magnetic field applied precisely along the medium
four-fold axis of a single molecule (i.e. $\theta=90^\circ$ and
$\phi=45^\circ$); these levels are labeled on the right-hand-side
of Fig.~1a according to an $\hat{S}_x$ basis, where $m_x$
(=~integer) is the projection of the total spin along the applied
field axis. The high-field EPR spectra are dominated by
transitions between adjacent levels, i.e. $\Delta m_x=\pm1$. If
one follows the EPR spectra to lower fields, into a region where
the Zeeman and axial terms in Eq.~1 become comparable ($g
\mu_B${\bf B}~$\sim|D|S$), one finds that the transitions may be
grouped into two categories: i) those between levels which evolve
from $m_z=\pm i$ ($i=$~integer) zero-field doublets, which we
label $\alpha$ (represented by blue sticks in Fig.~1a); and ii)
those between levels which evolve from adjacent zero-field
doublets, which we label $\beta$ (represented by red solid circles
in Fig.~1a). This distinction between $\alpha$ and $\beta$
resonances is based on a zero-field $\hat{S}_z$ basis, where $m_z$
is the projection of the total spin along the uniaxial direction
of the crystal $-$ the levels are labeled according to this
convention on the left-hand-side of Fig.~1a. Throughout the
remainder of the paper, we number all transitions according to the
absolute value of $m_x$ (high-field $\hat{S}_x$ representation)
associated with the level from which the transition was excited,
preceded by either $\alpha$ or $\beta$ (low-field $\hat{S}_z$
representation) in order to distinguish between the two categories
of resonances. Therefore, the highest field blue stick in Fig.~1a
corresponds to $\alpha10$, while the highest field red circle
corresponds to $\beta9$.\cite{note1}

In order to set the scene, we first review the status of earlier
single-crystal EPR studies. The observation most pertinent to the
present work is seen in Fig.~1b, which compares various calculations
of the $\alpha$ and $\beta$ transition frequencies with actual 8~K
EPR peak positions.\cite{HillPolyS9,HillPolyE} For these earlier
experiments, data were obtained with the field applied parallel to
the hard plane of a single-crystal sample to within an accuracy of
$\sim1.5^\circ$ ($\theta=90^\circ\pm1.5^\circ$). Within the hard
plane, a crude attempt was made to align the magnetic field along
the medium four-fold axis. Based on our more recent angle dependent
studies on accurately aligned single crystals (section~III and ref.
[\onlinecite{HillPRL03a}]), we estimate that such an alignment was
achieved to within an accuracy of approximately $10^\circ$ (i.e.
$\phi \sim 45^\circ \pm 10^\circ$). For the purposes of the
following discussion, any mis-alignment within the hard plane is not
important. In the absence of tilting, the calculated $\alpha$
transition frequencies decrease smoothly to zero (thick blue curves
in Fig.~1b), whereas the $\beta$ transition frequencies (thick red
curves in Fig.~1b) go through minima which are on the order of
90~GHz for the three highest field branches. The actual
$\beta$-transition data, on the other hand, deviate significantly
from these predictions, i.e. they do not exhibit a minimum
frequency, but instead follow monotonic curves to the lowest
frequencies investigated. The same trend has been noted for field
alignment along different directions within the hard
plane.\cite{HillPolyS9,HillPolyE,note0} For comparison, Fig.~2
displays representative experimental EPR spectra obtained with the
field approximately along the hard four-fold axis
($\sim\phi=0^\circ$), also within $\pm 1.5^\circ$ of the hard/medium
plane ($\theta=90^\circ\pm 1.5^\circ$);\cite{note0} note that the
$\beta$-resonances are again seen to the lowest frequencies studied
(44~GHz).

Following our more recent angle-dependent studies that provided
clear indications for a breakdown of the four-fold symmetry of
Mn$_{12}$-ac,\cite{HillPRL03a} and in light of the Cornia
model,\cite{CorniaPRL02} we conjecture that the anomalous behavior
of the $\beta$ transitions might be explained in terms of tilts of
the easy-axes of magnetization at a local scale. To better
illustrate this idea, we have included tilted field calculations
in Fig.~1b (thin curves: blue$-\alpha$, red$-\beta$) for a few of
the highest field transitions. Each tilted field curve belongs to
a family of curves which deviate successively (in $0.5^\circ$
increments) from one of the zero-tilt curves ($\theta=90^\circ$,
thick lines); the tilted field curves have been truncated at low
fields in order to simplify the figure. Tilting the field away
from the hard plane results in a field component parallel to the
easy axis. This ``longitudinal" field ($B_\parallel$) has the
effect of lifting the degeneracy of the of the $m_z=\pm i$
zero-field doublets in zeroth order.\cite{note2} Consequently,
instead of tending to zero-frequency, the $\alpha$ resonance
frequencies tend to successively larger offsets as the field is
tilted away from the hard plane, as seen by the thin blue curves
and indicated by the blue arrow (for $\alpha10$) in Fig.~1b. The
opposite is true for the $\beta$ transitions. Since the
longitudinal field splits the low-field $m_z=\pm i$ doublets, this
results in a reduction in the closest approach of levels in
adjacent doublets $-$ hence, to a reduction in the minimum
frequency of the $\beta$ transitions; again, this behavior is born
out by the thin red curves and indicated by the red arrow (for
$\beta9$) in Fig.~1b. Thus, tilting could explain the continuation
of the $\beta$ resonance data to frequencies well below the
theoretical minima given by the thick red curves
($\theta=90^\circ$, or $B_\parallel=0$) in Fig.~1b. However, the
observed trend cannot be ascribed to a simple mis-alignment of the
sample, because the $\alpha10$ (highest field blue squares) and
$\beta9$ (highest field red circles) resonances are both seen to
frequencies below 50~GHz (see also Fig.~2). The observation of
$\beta9$ at 45~GHz would imply a sample mis-alignment of
$2-3^\circ$, which is completely incompatible with the observed
behavior of the $\alpha10$ data which tracks the thick blue curve
($\theta=90^\circ$, or $B_\parallel=0$) in Fig.~1b; note that each
successive thin curve corresponds to an additional $0.5^\circ$ of
tilt away from the hard plane. Thus, it would appear that the data
in Figs.~1 and 2 reflect several molecular orientations.

A rigorous comparison between experiment and Eq.~1, in this
frequency/field range where the Zeeman and axial terms are
comparable, requires calculation of magnetic dipole matrix elements,
i.e. a full simulation of the EPR spectra. Figs.~3 and~4 show 15~K
EPR simulations, as a function of the polar angle $\theta$ between
the applied field and the easy axis of a single molecule, and for
two of the lowest frequencies used in our experiments (44~GHz and
62~GHz); the simulations are limited to the $\theta = 80^\circ$ to
$90^\circ$ range in $0.2^\circ$ increments. In Figs~3a and~4a, the
vertical scale represents microwave absorption while, in Figs~3b
and~4b, absorption is indicated by the darker shaded regions. The
simulations were generated using the software package
SIM;\cite{Weihe1,Weihe2} no rhombic ($E$) term was included at this
stage, and the simulations assume perfect alignment of all molecules
in the sample, i.e. no orientational averaging was employed. As
expected, the simulations confirm that only the $\alpha$ peaks are
observed for fields parallel to the hard plane ($\theta =
90^\circ$), for $f<90$~GHz. However, tilting the field away from the
hard plane leads to a very abrupt suppression of the magnetic dipole
matrix elements (EPR selection rules) for the $\alpha$ transitions
$-$ this behavior was not obvious from Fig.~1. In just $1^\circ$ of
rotation, $\alpha10$ vanishes completely at 44~GHz ($\sim1.5^\circ$
for 62~GHz). This is then followed by a range of almost $2^\circ$
where neither $\alpha10$ nor $\beta9$ are observed at the lowest
frequency of 44~GHz, i.e. $\beta9$ does not appear until the field
is tilted about $2.8^\circ$ away from the hard plane. Similar trends
are seen at both frequencies for $\alpha8$ and $\beta7$, etc..,
albeit over differing angle ranges. Remarkably, the simulations are
suggestive of a symmetry effect in the transverse-field ($B_\perp$)
EPR spectra, as a function of the longitudinal field component
($B_\parallel<B_\perp$). The dashed curves on the contour plots in
Figs.~3b and~4b indicate lines of constant $B_\parallel$,
demonstrating that all of the $\alpha$ resonances disappear (and all
of the $\beta$ resonances appear) at roughly the same longitudinal
field strengths. Subsequently, the $\beta$ resonances disappear and
the $\alpha^\prime$ resonances re-appear. This behavior is
reminiscent of the tunnel-splittings seen in the Fe$_8$ SMM as a
function of transverse field, for different static longitudinal
fields.\cite{WernsFe8} Indeed, closer inspection of Fig.~1b reveals
that the $\beta$ transition frequencies tend to zero at certain
values of $B_\parallel$, while the simulations in Figs.~3 and 4
suggest that the $\beta$ transitions matrix elements tend to zero as
well at these same angles. Thus, the analogy with the quenching of
the ground-state tunnel splittings in Fe$_8$ is significant, albeit
the degeneracies involve excited levels evolving from different
zero-field doublets in the present case. We shall not pursue this
analogy further here. However, we note that the $\beta$ series of
resonances is exactly what one would expect from a spin $S=9$ system
in a perpendicular field with the same Hamiltonian parameters as the
parent $S=10$ system;\cite{HillPolyS9,ZipseS9,PetukhovS9} meanwhile
$\alpha8^\prime$ corresponds to the first expected peak in a series
belonging to a spin $S=8$ system, and so on. Furthermore, all
resonances except for $\alpha10$ occur from excited levels. Thus,
the tilted field data have the appearance of spectra from excited
multiplets having successively lower total spin, as was originally
proposed by us as an explanation for the anomalous $\beta$
resonances.\cite{HillPolyS9}

An intensity analysis of the measured low-frequency ($<90$~GHz)
$\beta9$ peaks in Fig.~2 yields an (approximate) activation energy
consistent with an excitation from the $m_x = -9$ state within the
$S=10$ manifold. We should point out that, in order to precisely fit
the $\beta 9$ intensity, one must first fully comprehend its origin
(we save such an analysis until the end of this article). The
clearest indication that $\beta9$ likely occurs within the $S=10$
manifold can be seen by noting that the temperature dependencies of
the $\beta 9$ and $\alpha 8$ transitions in Fig.~2 are very similar,
suggesting that they both involve energy levels which are close in
energy. Thus, in the following sections, we {\em do not} consider
the possibility of excitations within excited state ($S\neq10$)
manifolds. However, the simultaneous observation of $\alpha10$ and
$\beta9$ down to a frequency of 44~GHz (Fig.~2) {\em is not}
consistent with the simulations in Fig.~3, which assume aligned
molecules. For this reason, we argue that the molecular easy-axes
are not perfectly aligned, i.e. we propose a distribution of
easy-axis alignments centered about the global four-fold axis of the
crystal. This would result in overlapping angle dependent features
(such as those in Figs.~3 and~4) from different parts of the
distribution, i.e. for a given field orientation one may observe
$\beta$ peaks from tilted (aligned) molecules together with $\alpha$
peaks due to aligned (tilted) molecules. As already discussed, the
origin for such a distribution can be understood in terms of solvent
disorder. Indeed, the calculations by Cornia et al. predict
easy-axis tilts of up to $0.5^\circ$.\cite{CorniaPRL02} The absence
of either $\alpha10$ or $\beta9$ over an approximately $2^\circ$
range in the simulation in Fig.~3 suggests that the tilt
distribution extends at least $\pm 1^\circ$, since both $\alpha10$
and $\beta9$ are observed in the 44~GHz spectra (Fig.~2).

\section{Experimental}

In order to confirm the above hypothesis, we have recently carried
out extremely precise angle-dependent measurements for $\theta$
rotations away from the hard plane ($\sim0.1^\circ$). In contrast
to earlier angle-dependent investigations within the hard plane
($\phi$ rotations),\cite{HillPRL03a} which were achieved using a
rotating split-pair magnet, the present study required fields
exceeding those available in the split-pair. Consequently, a
unique cavity was designed and constructed, enabling in-situ
rotation of the sample on the end-plate of the cylindrical cavity
with an angle step of $0.18^\circ$, and in fields of up to
45~tesla (at the National High Magnetic Field Laboratory). Details
concerning this cavity will be published
elsewhere,\cite{susumuRSI} and an account of our EPR spectrometer
is published in ref.~[\onlinecite{MolaRSI}]. The present
investigation was carried out in a standard 9~tesla
superconducting solenoid at the University of Florida. The sample
was positioned on the cavity end-plate so that the plane of
rotation coincided approximately with one of the four large flat
faces of the needle-shaped sample. Based on comparisons with
magnetic measurements,\cite{delBarco04} we believe that this plane
of rotation is inclined approximately $10^\circ$ with respect to a
plane containing the easy-axis and one of the hard four-fold axes,
i.e. we believe $\phi\sim -10^\circ$ in these investigations (see
Fig.~6 below).\cite{note0} Single crystal samples of
[Mn$_{12}$O$_{12}$(CH$_3$COO)$_{16}$(H$_2$O)$_4$]$\cdot$
2CH$_3$COOH$\cdot4$H$_2$O were grown using literature
methods.\cite{Lis} Magnetic field dependent absorption was
recorded at 15~K and at a frequency of 61.9~GHz, for field
orientations roughly $3.5^\circ$ either side of the hard plane
($\theta = 86.5^\circ\rightarrow93.5^\circ$), in $0.18^\circ$
increments. The experimental spectra displayed in Fig.~5 are
presented in several different ways in order to aid direct
comparison with Fig.~4. We note that the restricted bore size of
our high-field magnet forced a lower bound on the resonance
frequencies possible in the specially designed rotating
cylindrical cavity.\cite{susumuRSI} Thus, high-field measurements
below 61.9~GHz were not possible.

Immediately apparent from Fig.~5 is the fact that the $\alpha$ and
$\beta$ peaks overlap over a substantial angle range
($\sim2^\circ-3^\circ$) $\--$ particularly $\alpha10$, $\beta9$,
and $\alpha8~\--$ whereas this is not the case in Fig.~4, thus
providing direct confirmation for the tilt distribution. One also
observes a weak $\beta9$ peak in Fig.~5b for field alignment
precisely within the hard plane ($\theta=90^\circ$); in fact,
$\beta9$ is reduced below the $15\%$ level for only about
$0.5^\circ$ either side of the hard plane, in comparison to its
complete absence over a $90^\circ\pm 2^\circ$ range in Fig.~4b.
All of these facts suggest that the easy axes of some of the
molecules must be tilted by up to the order of $\pm 1.5^\circ$,
i.e. substantially greater than the tilts predicted by Cornia et
al.\cite{CorniaPRL02} Either that, or there exists a significant
distribution of internal transverse fields within the sample.
However, as previously discussed, this has been ruled out by
previous authors.\cite{Stamp} Indeed, internal fields are rather
weak,\cite{Parks,Kyungwha} and cannot explain the present
observations. Even for the worst possible sample shape,
depolarization effects could account for no more than a
$0.2^\circ$ distribution of field orientations at the applied
field strengths employed in these investigations. A more subtle,
yet significant, aspect of the data in Fig.~5 is the presence of
shoulders on the high-field sides of most of the EPR peaks (see
Fig.~5a). These shoulders have been emphasized in Fig.~5c via the
$50\%$ contour lines (this corresponds to $50\%$ of the maximum
absorption). Such shoulders have been noted previously by
us,\cite{HillPRL03a} and ascribed to the solvent-disorder-induced
rhombic anisotropy first proposed by Cornia et
al.\cite{CorniaPRL02} It is evident from Figs.~5a and~c that the
shoulder persists over a narrower angle range, when compared to
the main part of the EPR absorption peaks, i.e. suppression of the
shoulder relative to the main peak is quite abrupt, occurring
close to $\theta = 90\pm 1^\circ$ for $\alpha10$. The occurrence
of distinct regions in the 3D absorption plot (Fig.~5c), with
distinct angle dependencies, is suggestive of a discrete form of
disorder, as originally suggested in our earlier EPR
investigations.\cite{HillPRL03a,HillPRB02} As will become apparent
later in this article, the shoulder {\em is} attributable to a
finite rhombic term. However, the angle dependence cannot be
ascribed to a rhombic term alone $\--$ one must additionally
consider the combined effects of easy-axis tilting and rhombicity.
In doing this, one may understand the distinct angle dependence of
the main peaks and the shoulders (Fig.~5) as being due to
molecules tilted in distinct (orthogonal) planes.

\section{Discussion}

The origin of two separate angle dependencies finds a natural
explanation if one assumes that the easy-axis tilting and
rhombicity are connected (as predicted by Cornia), and that the
tilting is confined to directions determined by the principal axes
of the rhombic zero-field tensor. We now assume that two of the
Mn$_{12}$ variants originally discussed by Cornia et al. ($n=1$
and $n=3$)\cite{CorniaPRL02} have their easy axes tilted
significantly, and that these tilts are confined to two orthogonal
planes defined by the associated rhombic zero-field tensor and the
global four-fold ($z$-) axis. The $n=1$ and $n=3$ variants
comprise $50\%$ of the total number of molecules in Cornia's
model, and both species are predicted to have significant $E$
values ($1.6\times10^{-3}$~cm$^{-1}$).\cite{CorniaPRL02} Of this
$50\%$, half of the molecules ($25\%$ of the total $-$ the {\bf A}
molecules) are expected to have their hard two-fold axes aligned
along a single direction within the hard plane. Then, because
there exist four equivalent positions for the acetic acid of
crystallization related by a four-fold rotation about $z$, the
remaining half of the $n=1$ and $n=3$ species ($25\%$ of the total
$-$ the {\bf B} molecules) will have their hard axes aligned
$90^\circ$ away from the hard axes of the {\bf A} molecules, i.e.
parallel to the medium two-fold axes of the {\bf A} molecules.
This situation is depicted by the polar ($\phi$-) plot in the
lower panel of Fig.~6. The two-fold hard axes are denoted $HE$,
with the subscript {\bf A} or {\bf B} to distinguish between the
two sub-sets of molecule. The orientations of these axes relative
to the four-fold hard-axes ($HB^4_4$ at $\phi = 0^\circ$,
$90^\circ$, $180^\circ$, and $270^\circ$) is based on previous
angle-dependent EPR investigations for rotations within the hard
plane, where it was determined that the principal axes of the $E$
and $B^4_4$ tensors were mis-aligned by
$30^\circ$.\cite{HillPRL03a,note0} Next, we assume that the easy
axis tilts are also confined along these two distinct/discrete
directions, as indicated by the shaded regions in the lower panel
of Fig.~6. In other words, the well defined rhombic distortion and
molecular easy axis tilting are directly related. Based on the
ensuing analysis, we find that molecules tilted in one of the two
orthogonal planes have their $HE$ axes perpendicular to that
plane, as shown in the upper panel of Fig.~6. Thus, the {\bf A}
({\bf B}) molecules are tilted in a plane containing the $HE_B$
($HE_A$) axis.

It is now possible to see how one might expect two distinct angle
dependencies for different parts of the EPR spectrum, i.e. the
shoulders and the main peaks. The thick purple arrow along
$\phi=-10^\circ$ (and $170^\circ$) in the lower panel of Fig.~6
represents the best estimate ({\em vide infra}) of the plane of
rotation for the EPR spectra presented in Fig.~5. This plane is
inclined closer to $HE_A$ ($\phi=-30^\circ$ and $150^\circ$) than
to $HE_B$ ($\phi=60^\circ$ and $240^\circ$). Therefore, for fields
close to the hard plane ($\theta=90^\circ$), the rhombic term
($E\hat{O}^2_2$) associated with the $n=1$ and $n=3$ variants will
lead to different EPR peak-position-shifts for the {\bf A} and
{\bf B} molecules, thereby enabling one to resolve their
contributions to the EPR spectra for a sufficiently large $E$
value. In our experiments, the plane of rotation is closer to the
tilt-plane for the {\bf B} molecules than for the {\bf A}
molecules ($\Delta\phi = 20^\circ$ as opposed to $70^\circ$).
Thus, the projection of the tilt distribution onto the plane of
rotation is correspondingly greater for the {\bf B} molecules than
for the {\bf A} molecules. The magnitudes of these projections
scale as the $cosine$ of the angle ($\Delta\phi$) between the
field rotation plane and the corresponding tilt-plane. Indeed, as
explained later, it is on this basis that we deduced the
orientation of the plane of rotation relative to $HE_A$ and $HE_B$
for the experiments presented in Fig.~5. The molecules having the
greater projection of tilts onto the rotation plane will, thus, be
expected to exhibit the greater spread in their associated EPR
spectra away from the hard plane. In this scenario, one may
attribute the shoulders on the high-field sides of the main peaks
in Fig.~5 to the {\bf A} molecules in Fig.~6, since their tilt
distribution will have a narrower projection onto the rotation
plane (see projections onto field rotation plane represented by
blue and red dashed lines in Fig.~6). Furthermore, the fact that
the rotation plane is close to the hard axes of the {\bf A}
molecules is fully consistent with the appearance of the shoulder
on the high-field sides of the main peaks (assuming $E>0$).
Meanwhile, the contribution of the {\bf B} molecules is buried
within the main peak, as explained below.

Based on the geometry depicted in Fig.~6, we set out to simulate the
data in Fig.~5. The discussion above concerns only $50\%$ of the
possible variants in Cornia's model,\cite{CorniaPRL02} i.e. the two
dominant species possessing a significant $E$-term. Furthermore,
only half of these ($25\%$ of the total) are tilted in the plane of
rotation. For the sake of simplicity, we assume that the remaining
$50\%$ of the molecules ({\bf C}) do not possess a significant
$E$-value. This is a slight simplification of Cornia's model, since
it is known that the $n=2$ trans species (comprising $12.5\%$ of the
molecules) are predicted to have an $E$-value comparable to the
$n=1$ and $n=3$ variants.\cite{CorniaPRL02} However, one should
first recall that Cornia's calculations are approximate, and that
our previous EPR studies have indicated that the magnitudes of the
$E$-values may have been underestimated.\cite{HillPRL03a} Second,
introducing too many different species would result in an
over-parameterization of our model. The main purpose of the
following simulation is to demonstrate that one can explain
essentially all aspects of the single-crystal EPR spectra for
Mn$_{12}$-ac based on a simple model involving three just three
molecular species, together with discrete easy-axis tilting, in the
spirit of the original Cornia proposal. It is not our intent that
this model be viewed as definitive. Fig.~7 shows three gray-scale
contour plots representing the expected angle ($\theta$-) dependent
contributions to the EPR intensity for the three species {\bf A},
{\bf B}, and {\bf C}; they assume the same experimental conditions
as Figs.~4 and 5, i.e. T~$=$~15~K and a frequency of 62~GHz. The
simulations in Figs.~7a and~b assume that the {\bf A} and {\bf B}
molecules are tilted respectively along the $HE_A$ and $HE_B$
directions ($\phi = -30^\circ$ and $60^\circ$) i.e. the orientations
of the tilt planes are ``discrete." Meanwhile, we assume a
non-discrete distribution of the tilt angles along these two
directions, which cuts-off at $\theta = 90^\circ \pm 1.7^\circ$, as
indicated by the shaded region in the lower panel of Fig.~6. In
addition, we assume an $E$-value of $0.008$~cm$^{-1}$ for the {\bf
A} and {\bf B} species. The simulation in Fig.~7c assumes no
$E$-value. Consequently, the EPR intensity due to the {\bf C}
molecules contributes to the bulk of the central portion of the
summed EPR peaks (Fig.~8). We included a small random (i.e. not
confined to planes) distribution of tilts ($\pm 1^\circ$) for the
{\bf C} molecules as a means of taking into account possible
variations in the zero-field parameters/orientations associated with
the four Cornia variants ($n=0$, $n=4$, $n=2$ cis and trans) which
comprise this remaining $50\%$ of the molecules. It is the finite
$E$-value associated with the {\bf A} and {\bf B} species which
enables us to resolve the discrete nature of their tilting
directions. We cannot rule out similar properties for the remaining
species ({\bf C}), but their smaller $E$-values do not allow us to
observe such discrete behavior. Thus, the choice of random tilts for
the {\bf C} molecules is quite arbitrary. Nevertheless, this
distribution provides better overall qualitative agreement with the
observed out-of-plane angle dependence in Fig.~5.

Agreement between experiment (Fig.~5) and the subsequent summation
of the simulated spectra (Fig.~8) is rather sensitive to the precise
cut-off angle of $\theta = 90^\circ \pm 1.7^\circ$ for the tilt
distribution of the {\bf A} and {\bf B} molecules; however, it is
not so sensitive to the exact shape of the distribution. For
example, a discrete distribution with the {\bf A} and {\bf B}
molecules all having precisely $\theta = 90^\circ \pm 1.7^\circ$
tilts along $HE_A$ and $HE_B$ reproduces many aspects of the
experiments. However, such simulations contain many more fine
structures which are averaged out by considering a smoother
distribution such as the one employed in the Fig.~8 simulations.
Once again, we stress that this parameter set should not be viewed
as definitive. However, we emphasize that, in order to simulate the
main qualitative trends in the data (e.g. the distinct angle
dependencies of the main peaks and high-field shoulders), a discrete
tilt distribution is necessary. As discussed above, the simulations
are not so sensitive to the shape of the tilt distribution within
each tilt plane. Thus, it is not easy to make direct comparisons
with the distributions of tilts inferred by other
methods.\cite{Werns04} However, we note that only a small subset of
the molecules ($25\%$), corresponding to half of the lowest symmetry
Cornia variants,\cite{CorniaPRL02} are able to contribute to the
anomalous EPR intensity in our experiment. Of these molecules, we
can conclude that fewer than half ($\sim10\%$ of the total) are
tilted by more than $1^\circ$. While comparisons between the
gray-scale plots are not so sensitive to the $E$-value, comparisons
between simulated and actual experimental spectra are extremely
sensitive to $E$ (see Fig.~10 below); the value of
$0.008(2)$~cm$^{-1}$ gives the best agreement with a wide body of
single-crystal data collected over several
years.\cite{HillPRL03a,HillPRL98,HillPolyS9,HillPolyE,HillPRB02}
This $E$ value is also in excellent agreement with a recent first
principles calculation by Park et al.\cite{ParkPRB04}

As discussed above, the relative angle ($\Delta\phi$) between the
field rotation plane and the {\bf A} and {\bf B} tilt-planes ($HE_B$
and $HE_A$ respectively) was chosen so that the EPR peaks
corresponding to the {\bf A} and {\bf B} molecules (Figs.~7a and~b)
cover the same angle ranges as the high-field shoulders ($\theta
\sim 90^\circ \pm 1^\circ$ for $\alpha10$) and the central portions
of the peaks ($\theta \sim 90^\circ \pm 2^\circ$ for $\alpha10$)
respectively in Fig.~5. The best agreement was obtained with the
field rotation plane oriented $\Delta\phi = 20^\circ$ away from
$HE_A$ (Fig.~6). Thus, we estimate that the $HE_A$ and $HE_B$ axes
are oriented $\mp 20^\circ$ and $\pm 70^\circ$ with respect to the
sample faces, while the hard four-fold axes are roughly $\pm
10^\circ$ away (see Fig.~6), which also agrees with magnetic and
structural data.\cite{delBarco03,delBarco04} The simulated spectra
obtained according to the aforementioned procedure were then
normalized and summed in the ratio {\bf A}:{\bf B}:{\bf
C}~$\equiv$~$25\%:25\%:50\%$. The results of this summation are
displayed in Fig.~8. Agreement with the experimental data in Fig.~5
is quite impressive. First and foremost, the simulations including
tilts account for essentially all of the anomalous aspects of the
data which could not be explained without tilts (Fig.~4). For
example: the $\alpha10$ and $\beta9$ peaks overlap by about
$2^\circ$ between roughly $\theta = 90^\circ \pm0.5^\circ$ and
$90^\circ \pm2.5^\circ$; the $\beta9$ peak disappears (below the
$10\%$ level) over a roughly $\pm0.5^\circ$ range either side of
$\theta = 90^\circ$. There is also clear evidence for a shoulder on
the high-field sides of the main EPR peaks, although this shoulder
is more apparent in Fig.~5a than 5b. The overall widths and shapes
of the resonances are also reproduced fairly well, including the
broad low-field tail and the fairly abrupt decrease in intensity on
the high-field sides of the peaks (a 0.1~tesla Gaussian lineshape
was employed in the individual simulations in Fig.~7). We note from
Fig.~5 that an extremely weak $\beta 9$ signal persists even for
fields parallel to the hard plane. We were unable to reproduce this
behavior in the simulations while, at the same time, maintaining
good agreement with other aspects of the experimental data, e.g. the
angle range of overlap between the $\alpha$ and $\beta$ resonances.
We therefore speculate that this weak remnant signal may be due to
excitations within a higher lying $S=9$ state, as previously
proposed by us.\cite{HillPolyS9} However, more careful temperature
dependent studies will be required in order to establish the
location of this excited state relative to the $S=10$ ground state.
Indeed, recent measurements on a related high symmetry Mn$_{12}$
species (without significant easy axis tilting) have enabled
precisely such an analysis.\cite{PetukhovS9}

Having satisfactorily reproduced the spectra for out-of-plane
rotations (Figs.~5 and~8), we consider previously published
spectra obtained as a function of the field orientation $\phi$
within the hard plane.\cite{HillPRL03a} One of the puzzling
questions concerning these earlier measurements concerned the
observation of only a single satellite peak (shoulder) on the
high-field side of resonances for some field orientations. The
modulation of the widths of the EPR peaks was originally explained
in terms of the Cornia model.\cite{CorniaPRL02} So why does one
never see both low-field and high-field shoulders, corresponding
to the positive and negative EPR line-position-shifts induced by
the different signs of the rhombic term? We note that, for a field
along $HE_A$ ($HE_B$), one expects an upward (downward) shift in
field of the absorption due to the {\bf A} molecules, together
with a downward (upward) shift due to the {\bf B} molecules. The
apparent explanation is illustrated in Fig.~9, which displays the
evolution of absorption due to one of the tilted species (either
{\bf A} or {\bf B}) as a function of the field orientation (within
the hard plane) relative to the hard two-fold axis for that
species; note $\--$ this is not the same as the angle $\phi$. With
the field along the hard two-fold axis $HE$ ($\phi_E \equiv
0^\circ$ in Fig.~9), the resonance occurs at the highest field
position, while the projected distribution of tilts onto the
rotation plane is at its narrowest. Consequently, the EPR peak is
narrow. Conversely, with the field along the medium two-fold axis
($\phi_E \equiv 90^\circ$ in Fig.~9), the resonance occurs at the
lowest field position, and the projected distribution of tilts
onto the rotation plane is at its broadest, resulting in a broader
EPR peak. Conservation of the area under the peaks additionally
requires that the $\phi_E = 0^\circ$ peak be more intense at its
maximum than the $\phi_E = 90^\circ$ peak. The result is that the
broad low-field shoulder does not quite get resolved from the main
{\bf C} peak, whereas the sharp high-field peak is resolved for
several of the resonances $\--$ particularly $\alpha10$ and
$\alpha8$. In fact, the low-field $\phi_E = 90^\circ$ peak
contributes significantly to the slow decay of the $\alpha10$
resonance on its low-field side, thus accounting for another
previously unexplained feature of the Mn$_{12}$-ac single-crystal
EPR spectra. We note that recent experiments on deuterated samples
reveal shoulders on both the low and high field sides of the main
resonances.\cite{delBarco04}

Knowing the relative orientations of the principal axes associated
with the $E$ and $B^4_4$ tensors, we can simulate the full
hard-plane angle dependence published previously,\cite{HillPRL03a}
as shown in Figs.~10 to~12. We note that, while the misalignment of
the $E$ and $B^4_4$ tensors has previously been suggested from x-ray
structure measurements, the EPR studies in
ref.~[\onlinecite{HillPRL03a}] clearly illustrate that the magnetic
properties of Mn$_{12}$-ac reflect this misalignment. Fig.~10
displays the simulated spectra as a function of the field
orientation $\phi$ relative to one of the hard four-fold axes
($HB^4_4$). These spectra can be compared directly with Fig.~1 of
ref.~[\onlinecite{HillPRL03a}], albeit that the frequencies are
slightly different (50~GHz in ref.~[\onlinecite{HillPRL03a}] and
62~GHz in the present case). The previously published spectra do not
include the highest field $\alpha10$ resonance due to the
limitations of the split-pair magnet used in those investigations;
one should also be careful to note the different labeling scheme
used in this and earlier studies.\cite{note1} The splitting of
$\alpha8$ (highest field peak in ref.~[\onlinecite{HillPRL03a}]) is
reproduced very precisely, as are the amplitudes of the linewidths
(full-width-at-half-maximum $\--$ FWHM) variations with angle, as
displayed in Fig.~11 for peaks $\alpha2$ to $\alpha8$. Again, these
are the same four linewidths plotted in Fig.~2a of
ref.~[\onlinecite{HillPRL03a}]. The maximum linewidth or splitting
caused by the disorder-induced rhombic anisotropy is observed at
$\phi = \pm 60^\circ$ and $\mp 30^\circ$ relative to the $HB^4_4$
axes (see Fig.~6), and the widths are in fair agreement with the
$E$-value used in the present simulations. The orientations of the
linewidth maxima, which correspond to the hard and medium two-fold
axes ($HE_A$ and $HE_B$) are, therefore, oriented at $\pm 70^\circ$
and $\mp 20^\circ$ relative to the crystal faces. The $\pm$ and
$\mp$ symbols are used here to reflect the inversion symmetry of the
crystal. We note that the linewidth minima are somewhat deeper in
the simulation (Fig.~11), as compared to
experiment.\cite{HillPRL03a} This is most likely due to other
sources of line broadening (D-strain, g-strain, etc..) which were
not taken into consideration in our model. The main point of this
simulation is to once again show that the four-fold oscillation in
the linewidth/splitting is related to the $E$-strain, and that the
model employed in this study is in complete agreement with our
earlier investigations. The contour plot in Fig.~12 clearly
illustrates the four-fold variation in the line positions caused by
the intrinsic four-fold $B^4_4 \hat{O}^4_4$ transverse anisotropy.
The orientation of one the $HB^4_4$ axes, together with one of the
$HE$ axes, is indicated at the top of the figure, as is the
orientation of the rotation plane corresponding to the data in
Fig.~5. Overall, Fig.~12 is an excellent reproduction of the contour
plot in Fig.~1 of ref.~[\onlinecite{HillPRL03a}]. We remind the
reader that the linewidth/shape analysis was employed in
ref.~[\onlinecite{HillPRL03a}] to estimate the rhombic $E$-term
associated with the low-symmetry Mn$_{12}$ variants, and is
supported by independent magnetization studies.\cite{delBarco03}
Contrary to recent assertions,\cite{Werns04} the EPR linewidths are
not used in our determination of the easy-axis tilting, which has
been inferred entirely from the angle-dependent intensities of the
EPR spectra.

Finally, having reproduced the full angle dependence at a single
temperature of 15~K, we display in Fig.~13 a simulation of the
older temperature dependent data shown in Fig.~2. We found it
necessary to include a small mis-alignment ($1^\circ$, or
$\theta=89^\circ$) of the field away from the hard plane in order
to obtain the appropriate relative intensities of the $\alpha$ and
$\beta$ resonances, i.e. we suspect an approximately $1^\circ$
mis-alignment in the original experiment (we note that these
earlier studies were performed in a cryostat which did not allow
for in-situ alignment of the sample). Within the hard plane, the
simulation assumes that the field was applied along the $HB^4_4$
axis (i.e. $\phi = 0$). Agreement between the simulations and the
experiment is extremely good.

\section{Summary and conclusions}

Using angle-resolved single-crystal high-frequency EPR measurements,
we have clearly demonstrated that the molecular easy axes associated
with the widely studied Mn$_{12}$-ac single-molecule magnet are
tilted on a local scale. This tilting provides the much sought-after
source of the transverse fields necessary to explain the observation
of 'odd' resonant tunneling steps in hysteresis experiments, i.e.
the tilt distribution results in a transverse internal field
distribution, even when the field is applied precisely parallel to
the global easy-axis of the crystal. More importantly, on the basis
of the angle-dependence of fine structures observed in the EPR
spectra, we infer that the tilts are constrained along orthogonal
planes, i.e. there is a discrete aspect to the tilts, as was
previously suspected from studies which demonstrated the presence of
a locally varying rhombic anisotropy.\cite{HillPRL03a,delBarco03}
This behavior is completely reproducible in Mn$_{12}$-ac samples
prepared by different methods and by different
groups.\cite{delBarco04,HillChakov} Furthermore, our analysis is
insensitive to the intricacies associated with lineshape analyses,
which could be influenced by complex many-body effects, e.g.
magnetic spin-spin interactions.\cite{Werns04} Therefore, this study
provides compelling support for the recent predictions by Cornia et
al.,\cite{CorniaPRL02} who have suggested that a discrete form of
disorder associated with the acetic acid of crystallization provides
the dominant source of transverse anisotropy responsible for the
magnetic quantum tunneling in Mn$_{12}$-ac. Thus, one can rule out
suggestions that dislocations may be responsible for the symmetry
lowering at the local scale, since one would not expect the easy
axis tilting to be confined to discrete planes under such a
scenario.\cite{ChudnPRL01}

While our data agree with many qualitative aspects of the Cornia
model,\cite{CorniaPRL02} the magnitudes of the easy-axis tilting
angles (up to $1.7^\circ$), and the rhombic anisotropy associated
with the tilted molecules ($E$ up to $0.008$~cm$^{-1}$), are both
significantly larger than the predicted values (by a factor of about
5). These results should, therefore, be of significant theoretical
interest, forming a basis for more refined structure calculations
such as those recently published by Park et al.\cite{ParkPRB04} We
tentatively ascribe the observed tilting to the hydrogen-bonding
between the two acetic acid solvent molecules per Mn$_{12}$-ac. This
conjecture is supported by earlier observations that hydrogen
bonding must be important to the magnetization dynamics in the
Mn$_{12}$ cluster because deuteration leads to a significant slowing
down of the magnetization reversal dynamics as measured by
high-frequency ac susceptibility.\cite{Blinc} We note also that all
of the unusual solvent-induced anomalies reported in this paper for
Mn$_{12}$-ac are absent in the EPR spectra for a related
high-symmetry ($S_4$) Mn$_{12}$ complex
[Mn$_{12}$O$_{12}$(O$_{2}$CCH$_{2}$Br)$_{16}$(H$_{2}$O)$_{4}\cdot$4CH$_2$Cl$_2$]
which possesses a full compliment of four CH$_2$Cl$_2$ solvent
molecules per Mn$_{12}$, and none of the low symmetry hydrogen
bonding environments present in Cornia's model.\cite{PetukhovS9}


\bigskip

\section{Acknowledgements}

We thank E. del Barco, A. Kent, G. Christou, D. N. Hendrickson and
W. Wernsdorfer for useful discussion. This work was supported by the
NSF (DMR0103290 and DMR0239481). S. H. acknowledges Research
Corporation for financial support.

\section{References cited}

\clearpage

\noindent{{\bf Figure captions}}

\bigskip

FIG.~1. (a) Energy level diagram for Mn$_{12}$-ac with the field
applied in the hard plane ($\theta = 90^\circ$, $\phi =
45^\circ$); the crystal-field parameters used for the simulation
are given in the main text. The labels on the left hand side of
the figure correspond to the spin projection along $z$ (low-field
$m_z$ basis), while the labels on the right correspond to the spin
projection along the applied field direction (high-field $m_x$
basis). The vertical blue sticks illustrate the origin of the
$\alpha$ EPR transitions, while the filled red circles represent
the high-frequency $\beta$-transitions. (b) A comparison between
measured and calculated $\alpha$ and $\beta$ EPR transition
frequencies, as a function of the applied magnetic field strength.
The open blue squares ($\alpha$) and red circles ($\beta$)
represent experimental data obtained at 8~K. The thick curves
illustrate the expected behavior of the EPR transition frequencies
for precise field alignment within the hard plane
($\theta=90^\circ$), i.e. no tilting of the molecules. Note that,
for this case, the $\alpha$ frequencies decrease smoothly to zero
as the field tends to zero, whereas the $\beta$ frequencies go
through a minimum ($\sim90-95$~Ghz for the highest field
transitions) and then tend to a finite zero-field offset. The thin
curves illustrate the dependence of the $\alpha$ and $\beta$
transition frequencies upon tilting the field away from the hard
plane; each thin curve represents an additional $0.5^\circ$ of
tilt away from the hard plane ($\theta=90^\circ$), and the curves
have been truncated at low fields in order to simplify the figure.
The red arrow illustrates the depression of the minimum $\beta$
frequency, while the blue arrow illustrates the corresponding
increase in the minimum $\alpha$ frequency.

\bigskip

FIG.~2. Temperature and frequency dependence of the EPR spectra
obtained with the magnetic field applied approximately parallel to
the hard four-fold axis of the crystal;\cite{HillPolyS9,HillPolyE}
the frequencies and temperatures are indicated in the figure.
These experiments were performed using instrumentation that did
not allow for in-situ alignment of the sample. Based on subsequent
analysis, we estimate a slight ($\sim 1^\circ$) mis-alignment of
the field away from the hard plane (see also Fig.~13), i.e.
$\theta = 89^\circ$ and $\phi \sim 0^\circ$. The data clearly
demonstrate that both the $\alpha$ and $\beta$ resonances persist
to the lowest frequency (44.3~GHz), in apparent conflict with the
predictions of Fig.~1, thus suggesting that a significant fraction
of the molecules may be tilted with respect to the global easy
axis of the crystal (see text). The $\gamma$ resonance has been
previously attributed to a small fraction ($<5\%$) of Mn$_{12}$
molecules having lower symmetry (i.e. different crystal field
parameters) than the acetate.\cite{HillPolyS9}

\bigskip

FIG.~3. Simulations of 15~K, 44~GHz, EPR spectra as a function of
the field orientation $\theta$ (in $0.2^\circ$ increments)
relative to the easy axis of a single molecule ($\phi=0^\circ$).
In (a) the vertical scale represents absorption (arb. units)
while, in (b), absorption is indicated by the darker shaded
regions. The parameters used for these simulations are given in
the main text. In (b), the dashed curves represent constant values
of the longitudinal field $B_\parallel$. The key point to note is
the alternate appearance and disappearance of even $\alpha$ peaks
and odd $\beta$ peaks.

\bigskip

FIG.~4. Simulations of 15~K, 62~GHz, EPR spectra as a function of
the field orientation $\theta$ (in $0.2^\circ$ increments)
relative to the easy axis of a single molecule ($\phi=0^\circ$).
In (a) the vertical scale represents absorption (arb. units)
while, in (b), absorption is indicated by the darker shaded
regions. The parameters used for these simulations are given in
the main text. In (b), the dashed curves represent constant values
of the longitudinal field $B_\parallel$. The key point to note is
the alternate appearance and disappearance of even $\alpha$ peaks
and odd $\beta$ peaks.

\bigskip


\bigskip

FIG.~5. a) Experimental EPR spectra obtained at 15~K and 62~GHz,
as a function of the field orientation $\theta$ (in $0.18^\circ$
increments) relative to the global easy ($z$-) axis of a single
crystal sample; the red data represent $\theta=90^\circ$. These
experiments were performed using instrumentation that {\em did}
allow for in-situ rotation of the sample about 1
axis.\cite{susumuRSI} We estimate (see main text) that the
rotation plane was $10^\circ$ away from one of the hard $B^4_4$
axes within the hard ($xy$-) plane. To aid comparisons, the data
in b) are plotted in the same fashion as Fig.~4a, while the data
in c) are plotted in the same fashion as Fig.~4b. In (b) the
vertical scale represents absorption (arb. units). In (c)
absorption is indicated by the darker shaded regions; $15\%$ and
$50\%$ (of maximum absorption) contour lines have been included to
aid the discussion in the main text. No noise reduction/filtering
has been applied to the data. The main points to note are the
significant overlap of intensity due to the $\alpha$ and $\beta$
resonances, and the distinct shoulders on the high field sides of
the peaks which have been emphasized with the $50\%$ contour
lines.

\bigskip

FIG.~6. Schematic illustrating the discrete nature of the proposed
tilting of various Mn$_{12}$ species. Upper panel: illustration of
the tilt distribution for the {\bf A} molecules, wherein the tilts
are confined to the plane containing the crystallographic easy
($z$-) axis and the hard two-fold axis of the {\bf B} molecules
($HE_B \equiv$ medium two-fold axis of the {\bf A} molecules);
meanwhile, the hard two-fold axis of the {\bf A} molecules
($HE_A$) is perpendicular to the tilt plane; an orthogonal
situation exists for the {\bf B} molecules. Lower panel: a polar
plot ($\theta$ is radial and $\phi$ is azimuthal) illustrating the
orientations of the two tilt planes (shaded areas), and the
principal axes associated with the intrinsic four-fold anisotropy
($\hat{O}_4^4$) and disorder induced two-fold anisotropies
($\hat{O}_2^2$) associated with two of the Mn$_{12}$ variants
({\bf A} and {\bf B}); $HE_A$ and $HE_B$ refer respectively to the
hard two-fold axes for the {\bf A} and {\bf B} molecules; $HB_4^4$
and $MB_4^4$ refer respectively to the global hard and medium
four-fold axes; the purple arrow indicates the plane of rotation
for the experiments shown in Fig.~5, while the red and blue dashed
lines respectively indicate the extent of the projections of the
tilts of the {\bf A} and {\bf B} molecules onto the rotation
plane. Note the $\Delta \phi = 30^\circ$ offset between the
$\hat{O}_4^4$ and $\hat{O}_2^2$ tensors.

\bigskip

FIG.~7. Simulated contributions to the angle dependent EPR
spectrum for the {\bf A}, {\bf B} and {\bf C} molecules, for
$\theta$ rotations in a plane which is inclined $\Delta\phi =
20^\circ$ away from the tilt plane associated with the {\bf B}
molecules ($70^\circ$ away from the {\bf A} tilt plane, see also
Fig.~6). The temperature is 15~K and the frequency is 62~GHz; see
main text for a detailed explanation of these simulations. The
$E$-term associated with the {\bf A} and {\bf B} molecules splits
their contribution to the spectrum. The near coincidence of the
rotation plane and the hard axis of the {\bf A} molecules shifts
the contribution to the intensity to higher (lower) fields for the
{\bf A} ({\bf B}) molecules. Meanwhile, the near coincidence of
the rotation plane and the {\bf B} tilt plane (see Fig.~6)
projects almost the full tilt distribution for the {\bf B}
molecules, i.e. the absorption due to the {\bf B} molecules
extends over a wider angle range either side of the hard plane
($\theta = 90^\circ$) compared to either {\bf A} or {\bf C}, thus
accounting for the overlapping $\alpha$ and $\beta$ resonances in
Fig.~5. In contrast, the angle dependence of the EPR spectra
associated with the {\bf A} and {\bf C} molecules does not differ
greatly from the simulation in Fig.~4 (with no tilting).

\bigskip

FIG.~8. a) Angle dependent $25\%$:$25\%$:$50\%$ summations of the
normalized spectra for the {\bf A}, {\bf B} and {\bf C} molecules
from Fig.~7; each successive trace corresponds to a $0.2^\circ$
increment in $\theta$, and the red trace represents
$\theta=90^\circ$. In b), the same data are displayed in a 3D
gray-scale for direct comparisons with earlier figures; the $10\%$
and $50\%$ (of maximum absorption) contours are intended as guides
to the eye. These simulations are to be compared with the
experimental data in Fig.~5. In particular, the high-field
shoulder is reproduced (see also Fig.~10), as well as the
considerable angle overlap of the $\alpha$ and $\beta$ resonances.

\bigskip

FIG.~9. Simulated evolution of absorption due to either of the
tilted species ({\bf A} or {\bf B}) as a function of the field
orientation ($\phi_E$, within the hard plane) relative to the hard
two-fold axis for that species (note - this is not the same as the
angle $\phi$); the temperature is 15~K and the frequency is
62~GHz. With the field along the hard two-fold axis $HE$ ($\phi_E
\equiv 0^\circ$), the resonance occurs at the highest field
position, while the projected distribution of tilts onto the
rotation plane is at its narrowest. Consequently, the EPR peak is
narrow, forming a sharp shoulder on the high field side of the
summed spectra (see Figs.~8 and~10). Conversely, with the field
along the medium two-fold axis ($\phi_E \equiv 90^\circ$), the
resonance occurs at the lowest field position, and the projected
distribution of tilts onto the rotation plane is at its broadest,
resulting in a broader EPR peak, which contributes to the
low-field tail of the summed spectra.

\bigskip

FIG.~10. Simulated $\phi$-dependence of the 62~GHz, 15~K, EPR
spectra incorporating the $\Delta\phi = 30^\circ$ mis-alignment
between the intrinsic four-fold $\hat{O}^4_4$ tensor and the
disorder-induced two-fold $\hat{O}^2_2$ tensors for the {\bf A}
and {\bf B} molecules. These data compare favorably with
previously published data (see Fig.~1 of
ref.~[\onlinecite{HillPRL03a}]).

\bigskip

FIG.~11. Azimuthal angle ($\phi$-) dependence of the
full-widths-at-half-maximum (FWHM) for several of the $\alpha$
peaks in Fig.~10 ($\alpha 2$ to $\alpha 8$). This figure is to be
compared directly with Fig.~2a of ref.~[\onlinecite{HillPRL03a}].

\bigskip

FIG.~12. Grayscale contour plot corresponding to the data in
Fig.~11. The black dashed line indicates the orientation
corresponding to the data in Fig.~5. The orientations of one of
the $HE$ axes and one of the $HB_4^4$ axes are indicated at the
top of the figure. This figure is to be compared directly with the
contour plot Fig.~1 of ref.~[\onlinecite{HillPRL03a}] (note the
slightly different labeling used in these two
figures\cite{note1}); the white dashed lines in the present figure
correspond to the orientations of the two experimental curves in
Fig.~1 of ref.~[\onlinecite{HillPRL03a}].

\bigskip

FIG.~13. Full simulations of the temperature and frequency
dependent experimental data in Fig.~2 with the magnetic field
applied approximately parallel to the hard four-fold axis of the
crystal; the frequencies and temperatures are indicated in the
figure. In order to obtain the best agreement, a slight $1^\circ$
mis-alignment away from the hard plane was taken into account,
i.e. $\theta = 89^\circ$ and $\phi \sim 0^\circ$. Indeed, the
relative intensities of the $\alpha$ and $\beta$ resonances are
extremely sensitive to the angle $\theta$. This is easily
understood upon inspection of Fig.~1a, where it is seen that the
relative energy separations of the lowest lying states are
extremely sensitive to the field orientation. Therefore, the
corresponding transition probabilities, which depend on the
differences between Boltzmann factors, will also depend very
sensitively on the angle $\theta$.

\clearpage
\begin{figure}
\includegraphics[width=0.65\textwidth]{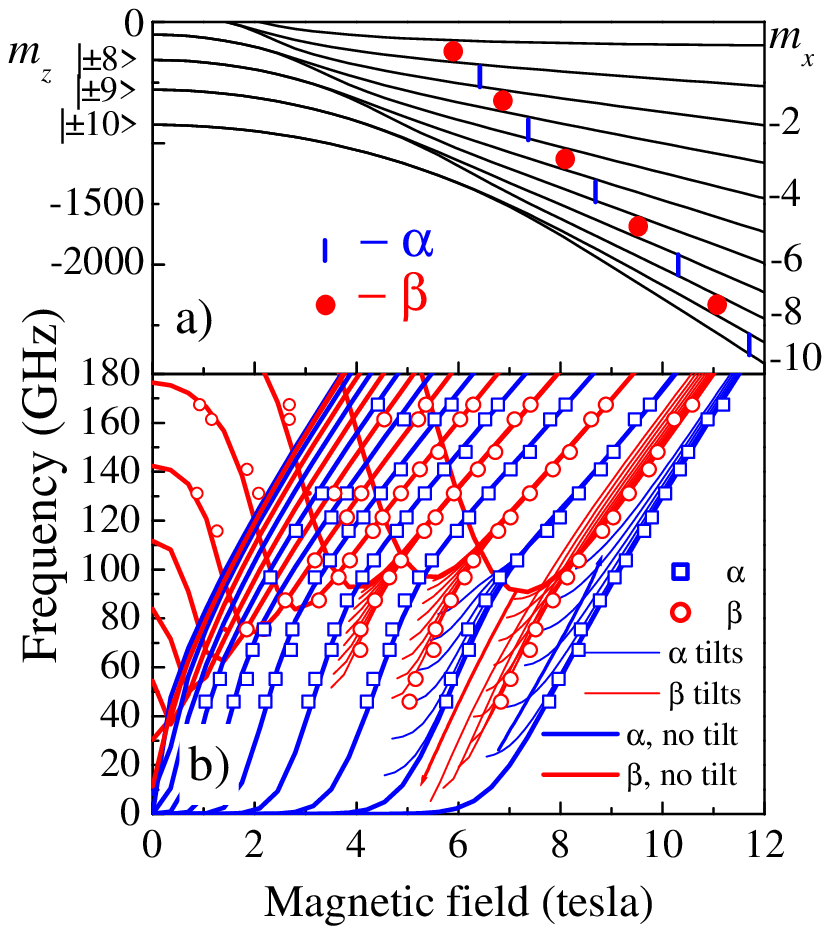}
\caption{\label{fig1} S. Hill {\em et al.}}
\end{figure}

\clearpage
\begin{figure}
\includegraphics[width=0.65\textwidth]{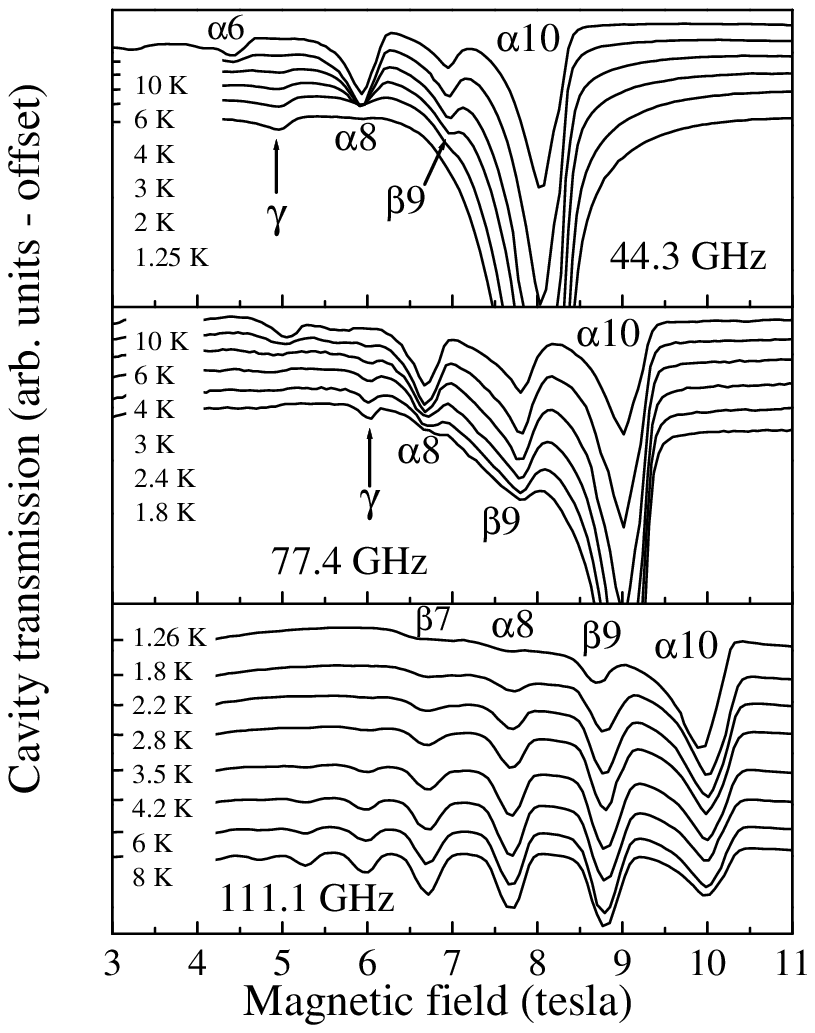}
\caption{\label{fig2} S. Hill {\em et al.}}
\end{figure}

\clearpage
\begin{figure}
\includegraphics[width=0.57\textwidth]{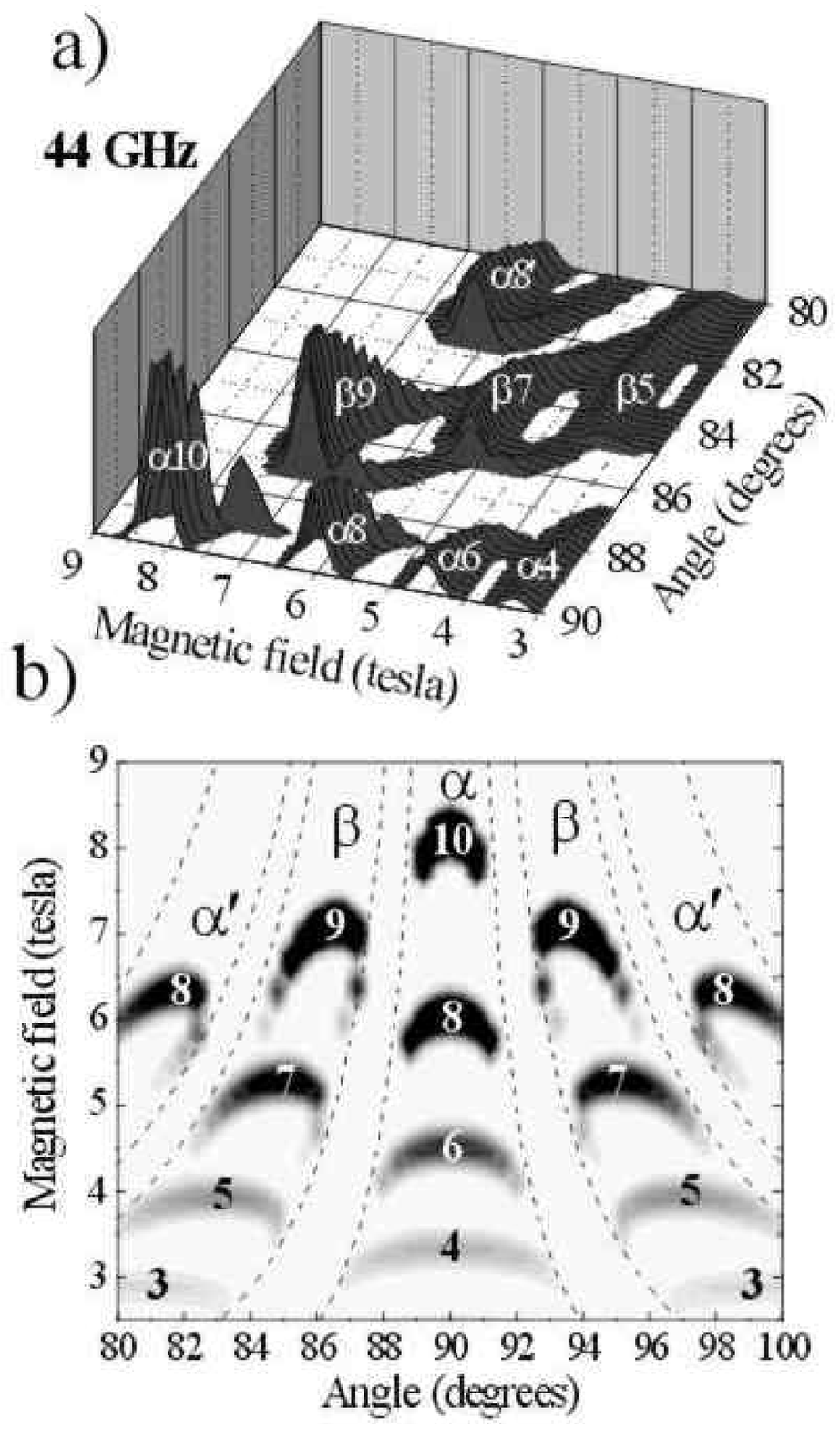}
\caption{\label{fig3} S. Hill {\em et al.}}
\end{figure}

\clearpage
\begin{figure}
\includegraphics[width=0.65\textwidth]{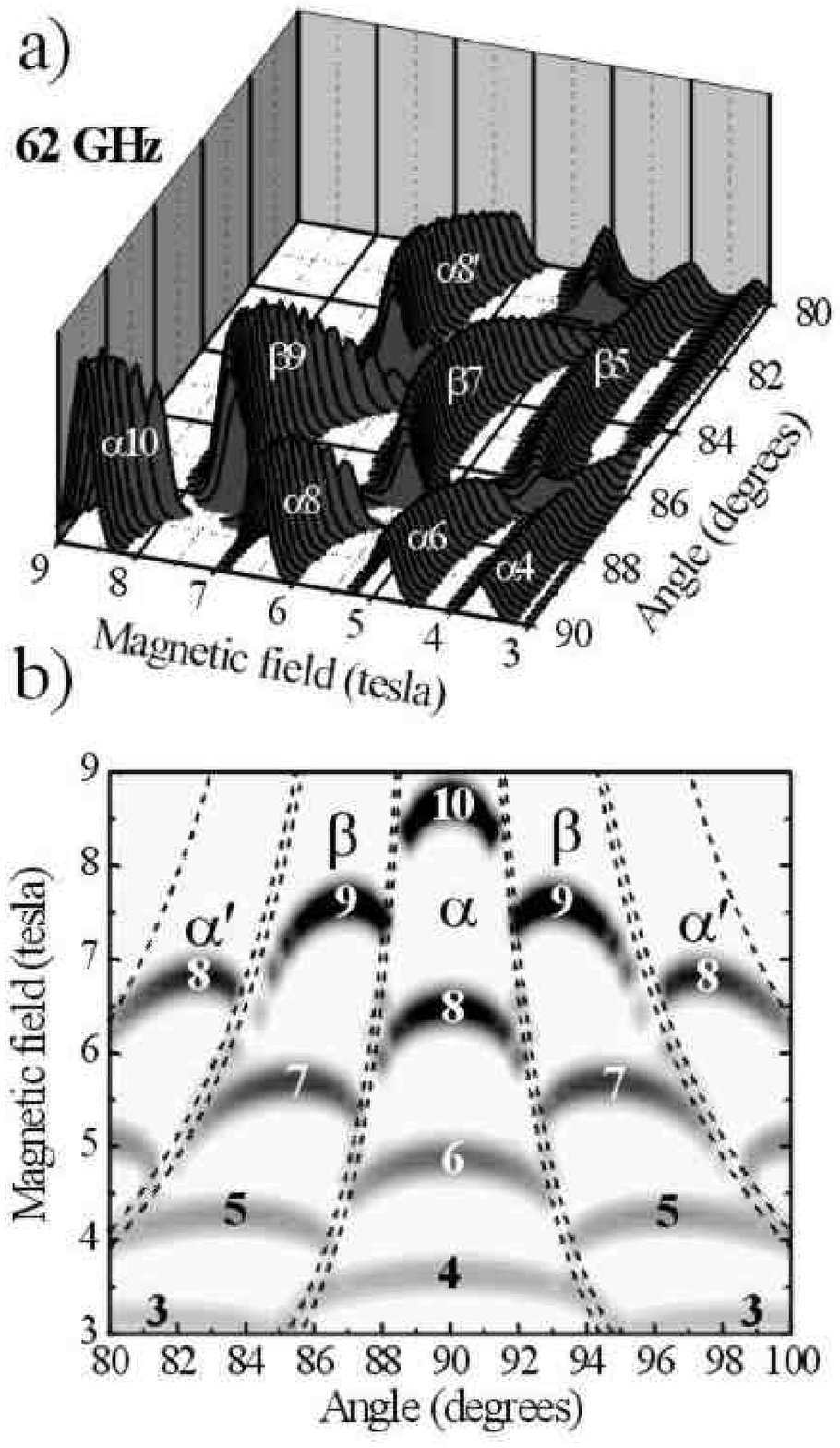}
\caption{\label{fig4} S. Hill {\em et al.}}
\end{figure}


\clearpage
\begin{figure}
\includegraphics[width=1.05\textwidth]{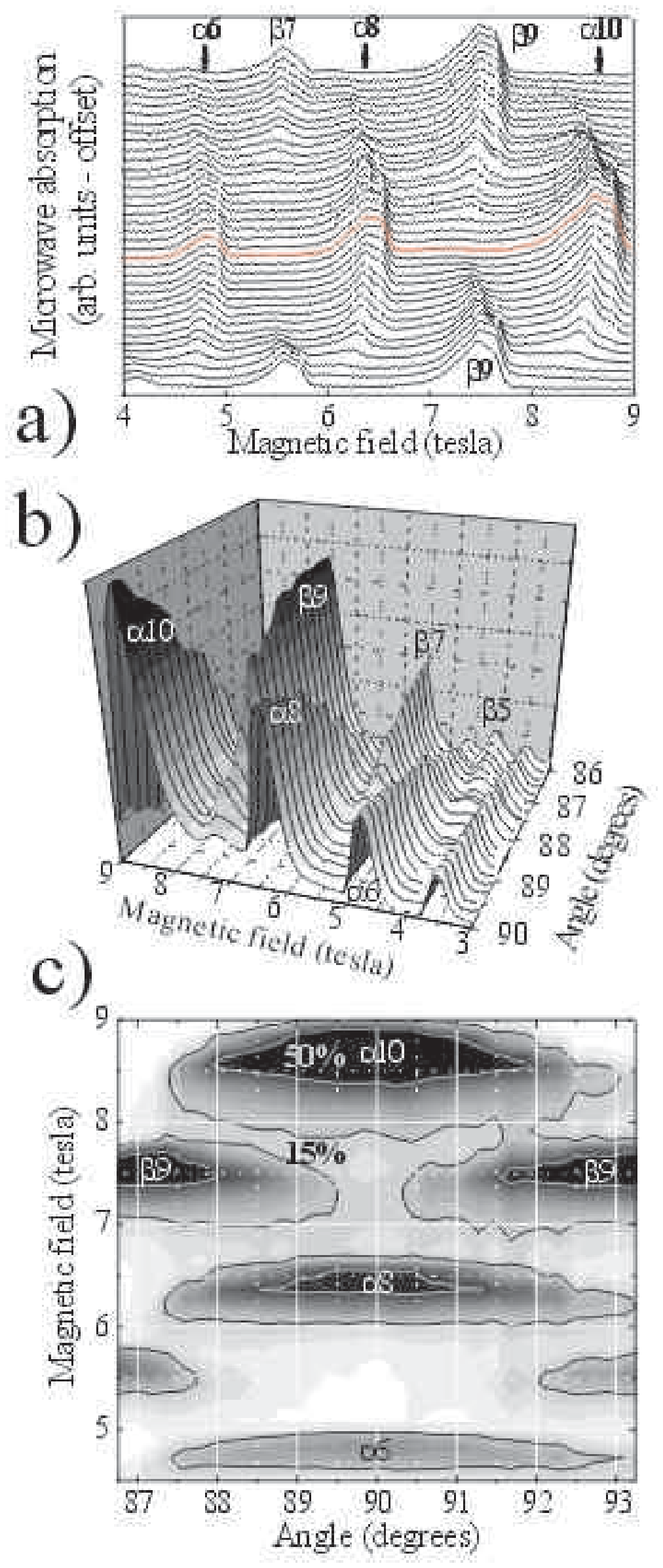}
\caption{\label{fig5} S. Hill {\em et al.}}
\end{figure}

\clearpage
\begin{figure}
\includegraphics[width=0.65\textwidth]{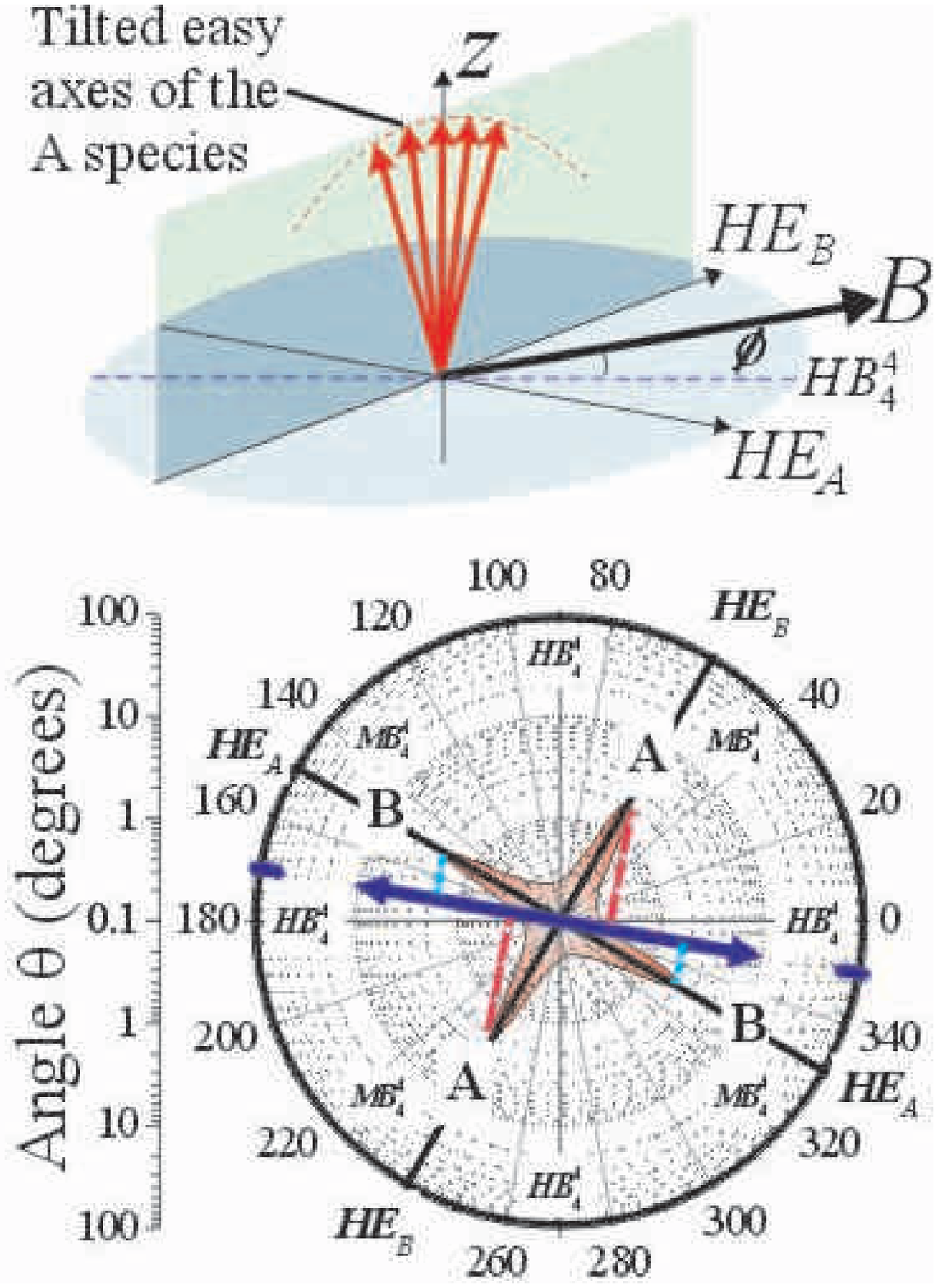}
\caption{\label{fig6} S. Hill {\em et al.}}
\end{figure}

\clearpage
\begin{figure}
\includegraphics[width=0.65\textwidth]{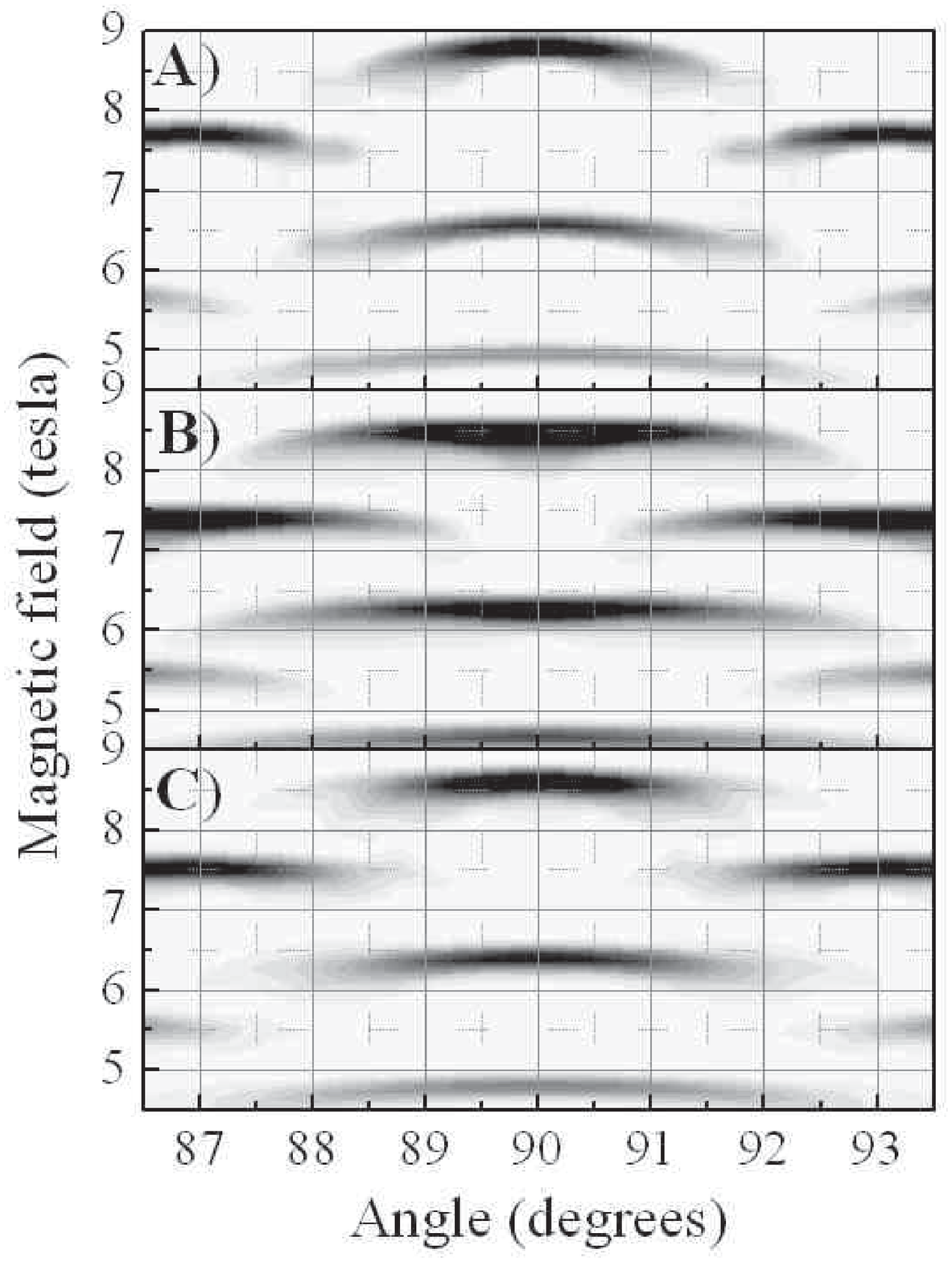}
\caption{\label{fig7} S. Hill {\em et al.}}
\end{figure}

\clearpage
\begin{figure}
\includegraphics[width=0.75\textwidth]{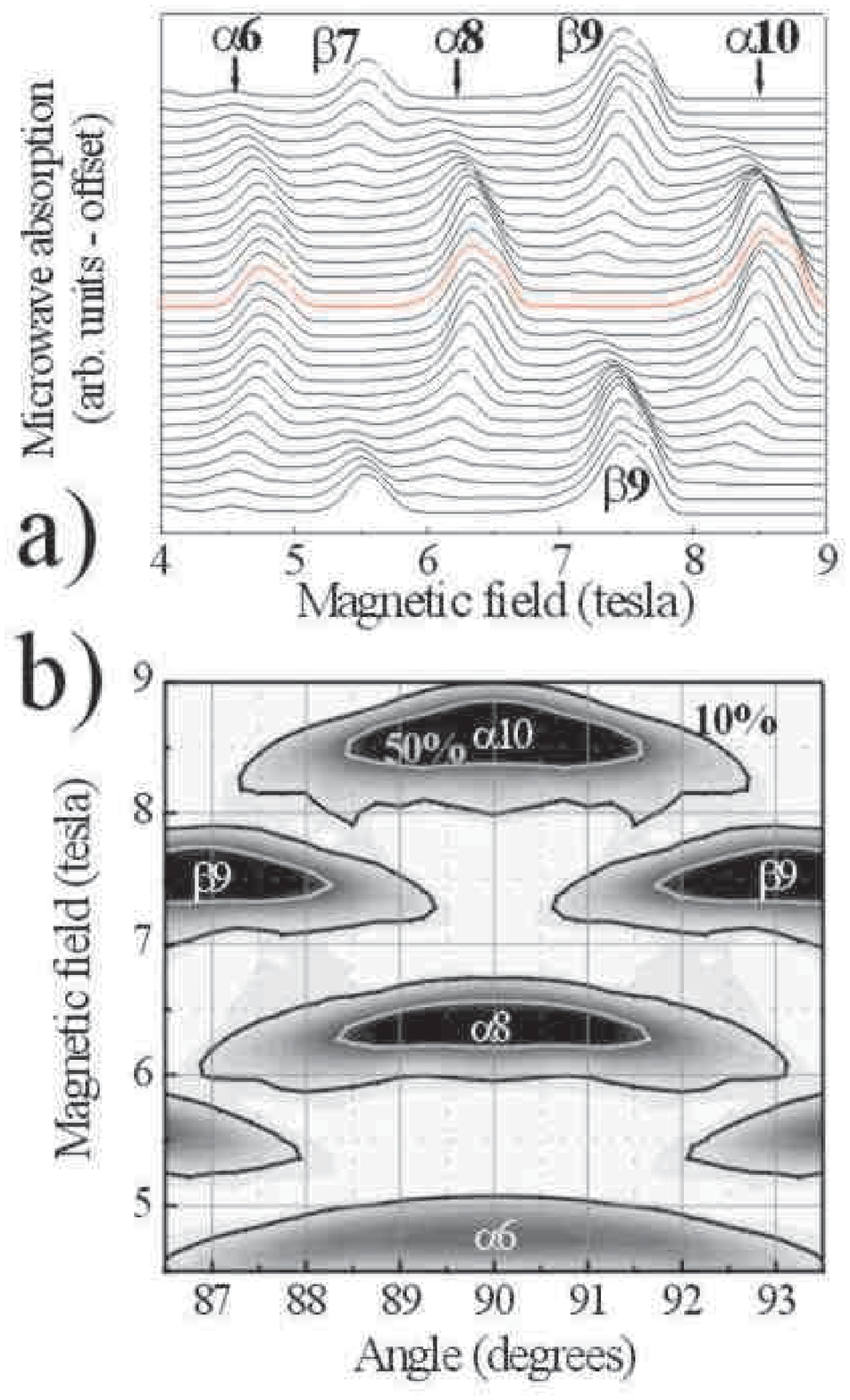}
\caption{\label{fig8} S. Hill {\em et al.}}
\end{figure}

\clearpage
\begin{figure}
\includegraphics[width=0.65\textwidth]{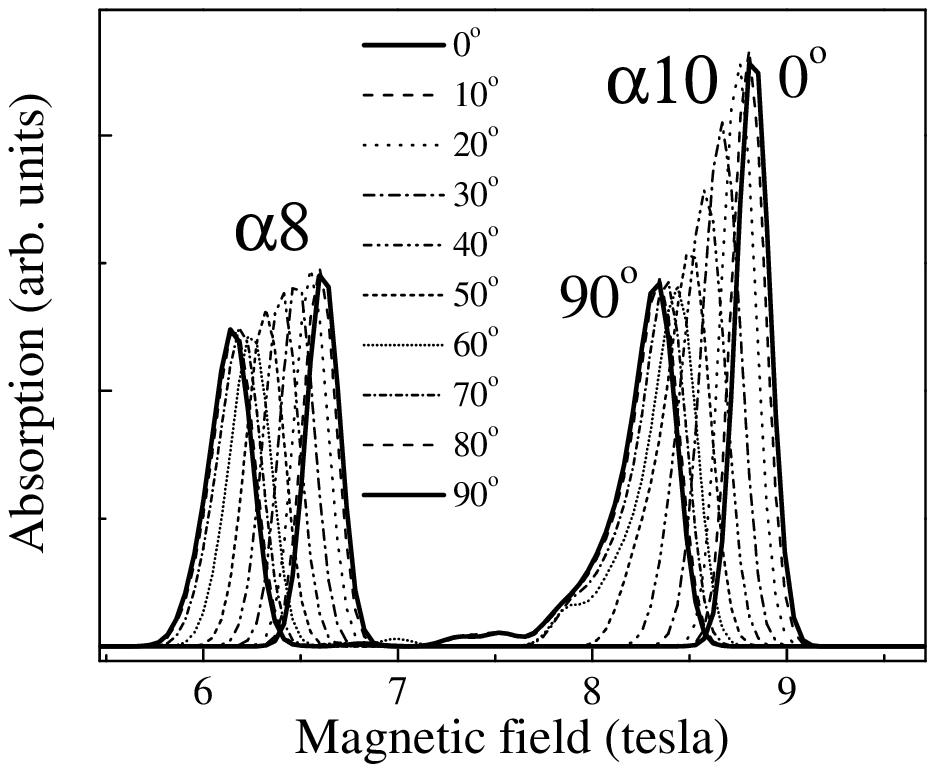}
\caption{\label{fig9} S. Hill {\em et al.}}
\end{figure}

\clearpage
\begin{figure}
\includegraphics[width=0.85\textwidth]{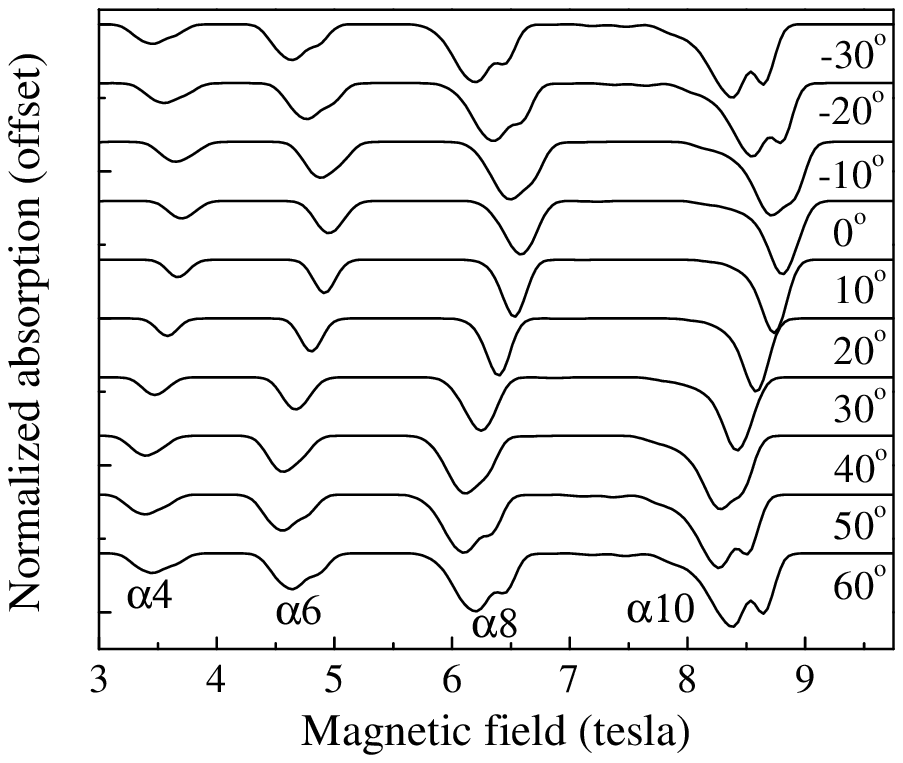}
\caption{\label{fig10} S. Hill {\em et al.}}
\end{figure}

\clearpage
\begin{figure}
\includegraphics[width=0.85\textwidth]{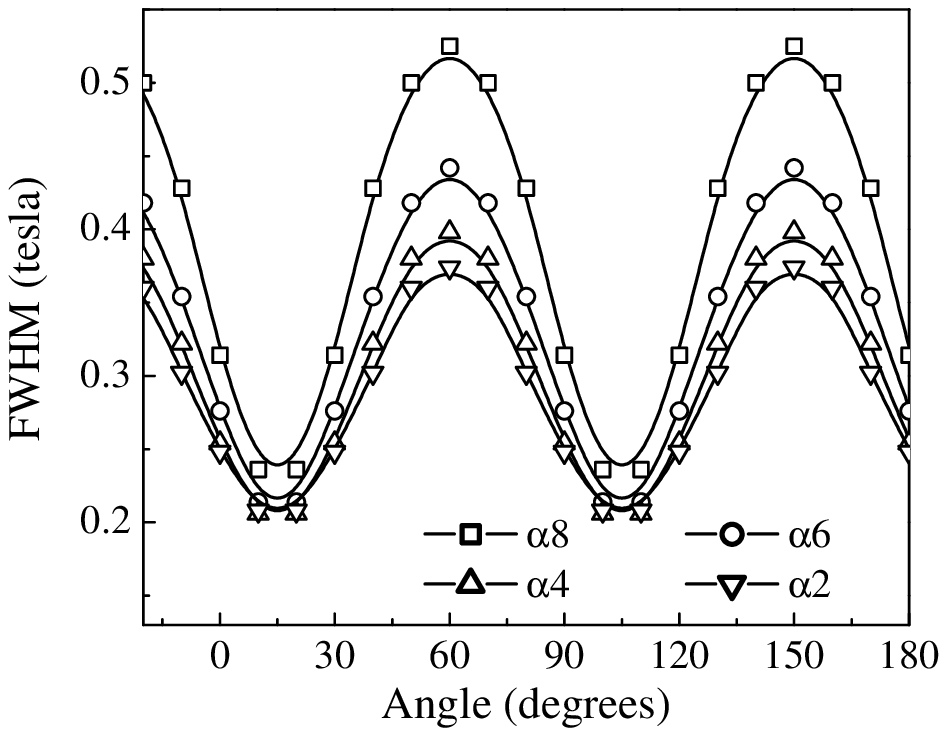}
\caption{\label{fig11} S. Hill {\em et al.}}
\end{figure}

\clearpage
\begin{figure}
\includegraphics[width=0.85\textwidth]{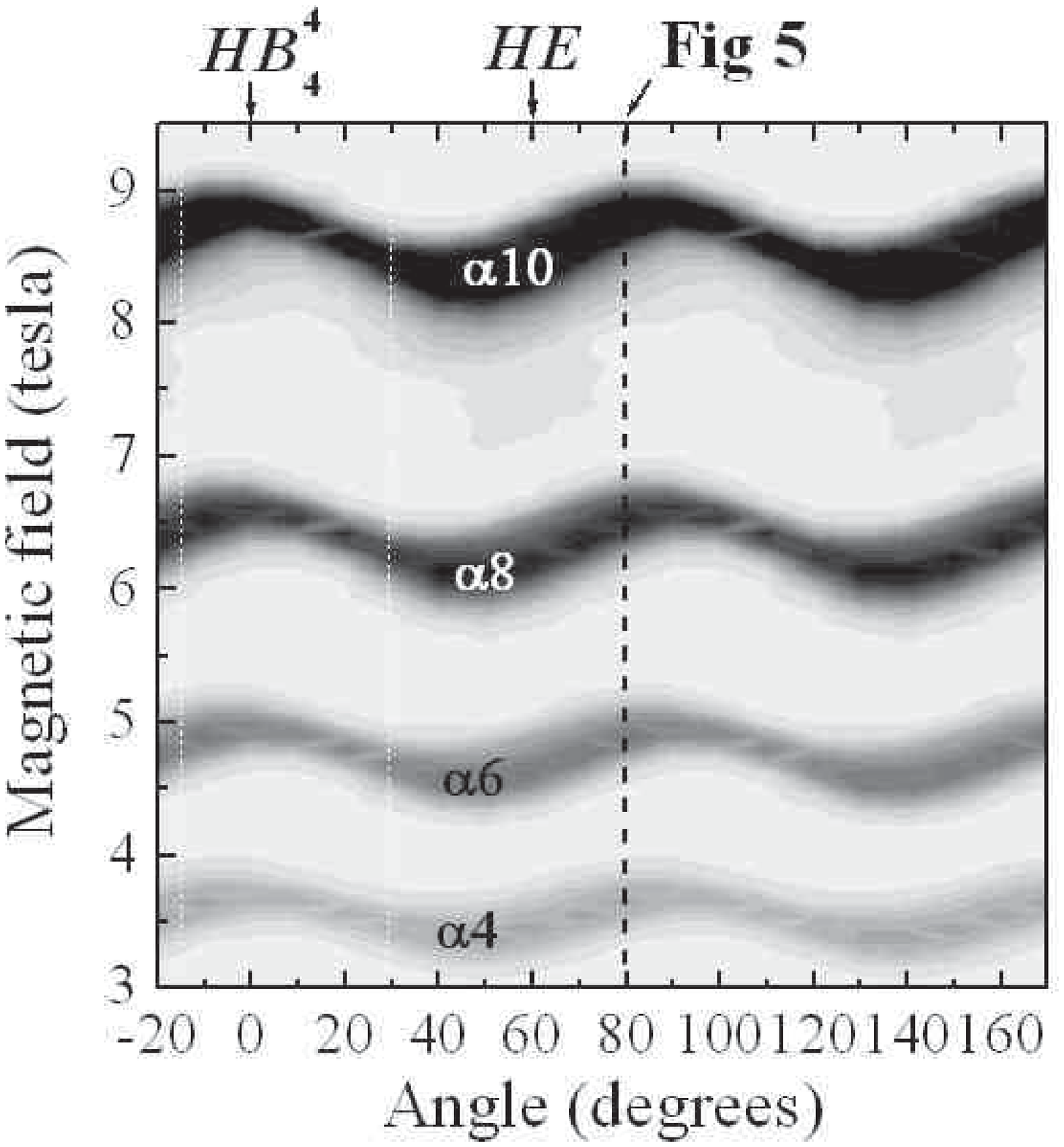}
\caption{\label{fig12} S. Hill {\em et al.}}
\end{figure}

\clearpage
\begin{figure}
\includegraphics[width=0.85\textwidth]{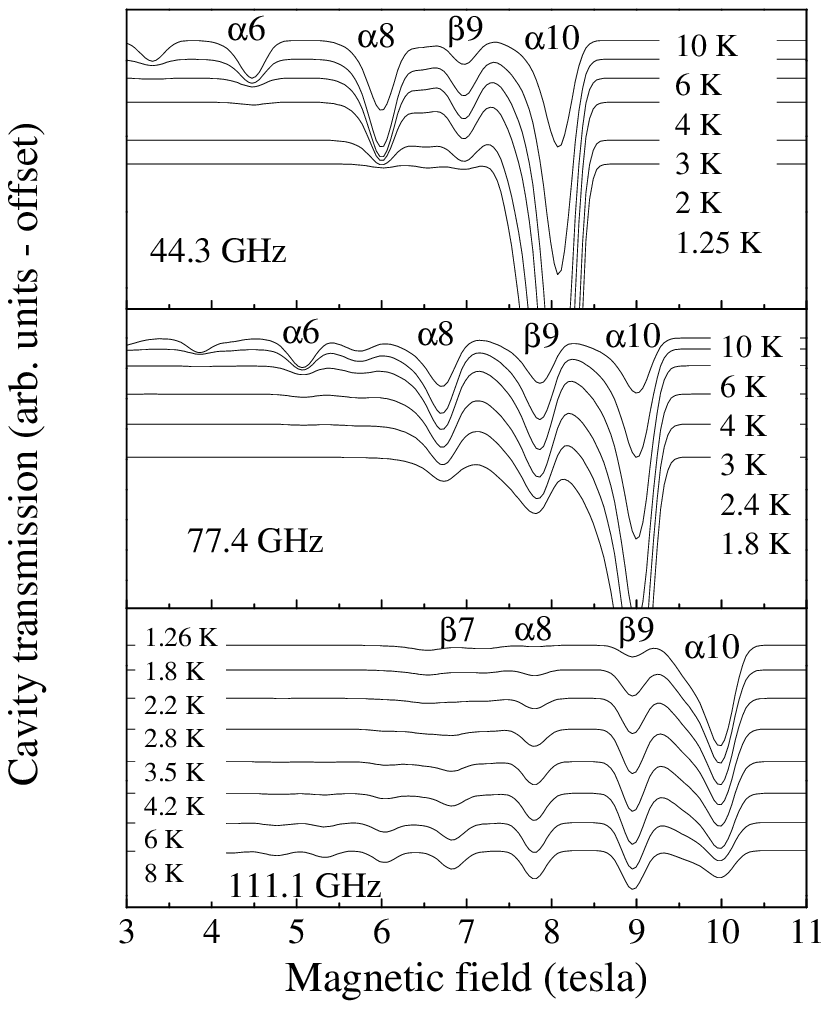}
\caption{\label{fig13} S. Hill {\em et al.}}
\end{figure}


\begin{thebibliography}{10}

\bibitem{Sessoli93a}
R. Sessoli, H.-L. Tsai, A. R. Schake, S. Wang, J. B. Vincent, K.
Folting, D. Gatteschi, G. Christou and D. N. Hendrickson, J. Am.
Chem. Soc. {\bf 115}, 1804 (1993)

\bibitem{Sessoli93b}
R. Sessoli, D. Gatteschi, A. Caneschi and M. Novak, Nature {\bf
365}, 141 (1993).

\bibitem{Sessoli95}
R. Sessoli, Mol. Cryst. Liq. Cryst. {\bf 274}, 145 (1995).

\bibitem{Novak95}
M. A. Novak, R. Sessoli, A. Caneschi and D. Gatteschi, J. Magn.
Magn. Mater. {\bf 146}, 211 (1995).

\bibitem{Friedman96}
J.~R. Friedman, M. P. Sarachik, J. Tejada and R. Ziolo, Phys. Rev.
Lett. {\bf 76},  3830 (1996).

\bibitem{Thomas96}
L. Thomas, F. Lionti, R. Ballou, D. Gatteschi, R. Sessoli and B.
Barbara, Nature {\bf 383},  145  (1996).

\bibitem{MRS00}
G. Christou, D. Gatteschi, D. Hendrickson, and R. Sessoli, MRS
Bulletin {\bf 25},  66  (2000).

\bibitem{Angewandte03}
D. Gatteschi and R. Sessoli, Angew. Chem. {\bf 42},  268  (2003).

\bibitem{LossNature01}
M.~N. Leuenberger and D. Loss, Nature {\bf 410},  789  (2001).

\bibitem{HillScience}
S. Hill, R. S. Edwards, N. Aliaga-Alcalde and G. Christou, Science
{\bf 302}, 1015 (2003).

\bibitem{Perenboom}
J. A. A. J. Perenboom, J. S. Brooks, S. Hill, T. Hathaway and N.
S. Dalal, Phys. Rev. B {\bf 58}, 330 (1998).

\bibitem{ChudnPRL01}
E.~M. Chudnovsky and D.~A. Garanin, Phys. Rev. Lett. {\bf 87},
187203  (2001).

\bibitem{CorniaPRL02}
A. Cornia, R. Sessoli, L. Sorace, D. Gatteschi, A. L. Barra and C.
Daiguebonne, Phys. Rev. Lett. {\bf 89},  257201 (2002).

\bibitem{HillPRL03a}
S. Hill, R. S. Edwards, S. I. Jones, N. S. Dalal and J. M. North,
Phys. Rev. Lett. {\bf 90}, 217204 (2003).

\bibitem{delBarco03}
E. del Barco, A. D. Kent, E. M. Rumberger, D. N. Hendrickson and
G. Christou, Phys. Rev. Lett. {\bf 91}, 047203 (2003).

\bibitem{WernsEPL99}
W. Wernsdorfer, R. Sessoli and D. Gatteschi, Europhysics Letters
{\bf 47}, 254 (1999).

\bibitem{Stamp}
N. V. Prokof'ev and P. C. E. Stamp, Phys. Rev. Lett. {\bf 80},
5794 (1998).

\bibitem{MertesPRL}
K. M. Mertes, Y. Suzuki, M. P. Sarachik, Y. Paltiel, H. Shtrikman,
E. Zeldov, E. Rumberger, D. N. Hendrickson and G. Christou, Phys.
Rev. Lett. {\bf 87}. 227205 (2001).

\bibitem{delBarcoEPL}
E. del Barco, A. D. Kent, E. M. Rumberger, D. N. Hendrickson and
G. Christou, Europhysics Letters {\bf 60}, 768 (2002).

\bibitem{Amigo}
R. Amigo, E. del Barco, Ll. Casas, E. Molins, J. Tejada, I. B.
Rutel, B. Mommouton, N. Dalal and J. Brooks, Phys. Rev. B {\bf
65}, 172403 (2002).

\bibitem{delBarco04b}
E. del Barco, A. D. Kent, N. E. Chakov, L. N. Zakharov, A. L.
Rheingold, D. N. Hendrickson and G. Christou, Phys. Rev. B {\bf
69}, 020411(R) (2004).

\bibitem{HillPRL98}
S. Hill, J. A. A. J. Perenboom, N. S. Dalal, T. Hathaway, T.
Stalcup and J. S. Brooks, Phys. Rev. Lett. {\bf 80},  2453 (1998).

\bibitem{delBarco04}
E. del Barco, A. D. Kent, S. Hill, J. M. North, N. S. Dalal, E. M.
Rumberger, D. N. Hendrickson, N. Chakov, G. Christou (unpublished);
also arXiv/cond-mat/0404390.

\bibitem{HillChakov}
S. Hill, N. E. Chakov, N. S. Dalal, G. Christou (unpublished).

\bibitem{Werns04}
W. Wernsdorfer, N. E. Chakov, G. Christou, arXix/cond-mat/0405565
(unpublished).

\bibitem{PetukhovS9}
K. Petukhov, S. Hill, N. E. Chakov, G. Christou, K. Abboud, Phys.
Rev. B (in-press, 2004); also arXix/cond-mat/0403435.

\bibitem{Barra}
A. L. Barra, D. Gatteschi, R. Sessoli, Phys. Rev. B {\bf 56}, 8192
(1997).

\bibitem{MirebeauPRL99}
I. Mirebeau, M. Hennion, H. Casalta, H. Andres, H. U. Gudel, A. V.
Irodova, and A. Caneschi, Phys. Rev. Lett. {\bf 83}, 628 (1999).

\bibitem{Mukhin}
A. Mukhin, V. D. Travkin, A. K. Zvezdin, S. P. Lebedev, A.
Caneschi, and D. Gatteschi, Europhys. Lett. {\bf 44}, 778 (1998).

\bibitem{HillPolyS9}
S. Hill, R. S. Edwards, J. M. North, S. Maccagnano and N. S.
Dalal, Polyhedron {\bf 22}, 1897 (2003).


\bibitem{HillPolyE}
S. Hill, R. S. Edwards, J. M. North, K. Park and N. S. Dalal,
Polyhedron {\bf 22}, 1889 (2003).

\bibitem{note0}
The sign of $B^4_4$ dictates the hard and easy directions of
magnetization, which are separated by $45^\circ$ from each other
in the $xy$-plane; the hard directions are along $x$ and $y$ for
$B^4_4>0$, and at $45^\circ$ to $x$ and $y$ for $B^4_4<0$. In a
recent investigation, we estimated that the hard axes are
approximately aligned with the edges of the square cross-section
of a typical Mn$_{12}$-ac single crystal sample, with the medium
axes oriented approximately accross the diagonals. Based on more
comprehensive studies, including comparisons between the present
work and recent magnetic measurements,\cite{delBarco03} we now
believe that the four-fold axes are shifted approximately
$10^\circ \pm 5^\circ$ away from the crystal faces, i.e. at
$10^\circ$, $55^\circ$, $90^\circ$, $145^\circ$, etc. This does
not invalidate any of the conclusions of our earlier studies,
particularly the relative orientations of the $E$ and $B_4^4$
tensors. A full account of the symmetries associated with the MQT
in Mn$_{12}$-ac will be presented elsewhere.\cite{delBarco04}

\bibitem{note1}
We note that this numbering convention differs slightly from the
scheme used in earlier publications. We previously used all
integers when numbering the $\alpha$ and $\beta$ resonances. This
scheme ignores the obvious symmetries associated with the various
transitions. Therefore, we now use only even integers for $\alpha$
resonances, and odd integers for $\beta$ resonances, i.e.
$\alpha9$ under the old notation is now $\alpha8$, $\beta10$ is
$\beta9$, $\alpha8$ is $\alpha6$, and so on.\cite{HillPRL98}

\bibitem{note2}
In contrast to $B_\parallel$, the ``transverse" field ($B_\perp$)
acts on the low-field $m_z=\pm i$ doublets only in high orders of
perturbation theory (the order depends explicitly on the nature of
other transverse terms in $\hat{H}$). At high fields ($g
\mu_B${\bf B}~$> |D|S$), the Zeeman term operates only in zeroth
order, since it defines the appropriate basis functions.

\bibitem{Weihe1}
Written by H. Weihe, University of Copenhagen,
http://sophus.kiku.dk/software/epr/epr.html.

\bibitem{Weihe2}
C. J. H. Jacobsen, E. Pedersen, J. Villadsen, H. Weihe, Inorg.
Chem. {\bf 32}, 1216-1221 (1993).

\bibitem{WernsFe8}
W. Wernsdorfer and R. Sessoli, Science {\bf 284}, 133 (1999).

\bibitem{ZipseS9}
D. Zipse, J. M. North, N. S. Dalal, S. Hill, R. S. Edwards, Phys.
Rev. B {\bf 68}, 184408 (2003).

\bibitem{susumuRSI}
S. Takahashi and S. Hill, submitted.

\bibitem{Parks}
B. Parks, J. Loomis, E. Rumberger, D. N. Hendrickson, G.
Christou, Phys. Rev. B {\bf 64}, 184426 (2001).

\bibitem{Kyungwha}
K. Park, M. A. Novotny, N. S. Dalal, S. Hill, P. A. Rikvold, Phys.
Rev. B {\bf 66}, 144409 (2002).

\bibitem{MolaRSI}
M. Mola, S. Hill, P. Goy and M. Gross, Rev. Sci. Inst. {\bf 71},
186 (2000).

\bibitem{Lis}
T. Lis, Acta Crystallogr. {\bf B36}, 2042 (1980).

\bibitem{HillPRB02}
S. Hill, S. Maccagnano, K. Park, R. M. Achey, J. M. North and N.
S. Dalal, Phys. Rev. B {\bf 65}, 224410 (2002).

\bibitem{ParkPRB04}
K. Park, T. Baruah, N. Bernstein, and M. R. Pederson, Phys. Rev. B
{\bf 69}, 144426 (2004).

\bibitem{Blinc}
R. Blinc, B. Zalar, A. Gregorovic, D. Arcon, Z. Kutnjak, C.
Filipic, A. Levstik, R.M. Achey and N.S. Dalal, Phys. Rev. B {\bf
67}, 094401 (2003).

\end{thebibliography}
\end{document}